\documentclass[12pt]{article}

\usepackage{graphicx, amsmath, amssymb, amsthm, multicol, multirow, bbold, float, subcaption, verbatim, enumitem, rotating, algorithmic}
\usepackage[utf8x]{inputenc}
\usepackage{epstopdf}

\usepackage[boxed]{algorithm2e}

\usepackage{thmtools}
\usepackage{thm-restate}

\usepackage{cite}

\usepackage{color}   %May be necessary if you want to color links
\usepackage{hyperref}
\hypersetup{
	hidelinks,
    linktoc=all     %set to all if you want both sections and subsections linked
}

\newtheorem{theorem}{Theorem}[section]
\newtheorem{definition}{Definition}%[section]
\newtheorem{lemma}[theorem]{Lemma}%[section]

\theoremstyle{definition}
\newtheorem{remark}[theorem]{Remark}
\newtheorem{example}[theorem]{Example}
\newtheorem{assumption}[theorem]{Assumption}

\numberwithin{equation}{section}

\newcommand{\cl}{\operatorname{cl}}
\newcommand{\interior}{\operatorname{int}}
\renewcommand{\succ}{\operatorname{succ}}

\newcommand{\trans}[1]{#1^{\mathsf{T}}}

\newcommand{\N}{\mathbb{N}}

\begin{document}
\title{Time consistency of the mean-risk problem}
\author{Gabriela Kováčová\thanks{Vienna University of Economics and Business, Institute for Statistics and Mathematics, Vienna A-1020, AUT, gabriela.kovacova@wu.ac.at and birgit.rudloff@wu.ac.at.} \and Birgit Rudloff\footnotemark[1] %\thanks{Vienna University of Economics and Business, Institute for Statistics and Mathematics, Vienna A-1020, AUT, birgit.rudloff@wu.ac.at.}
}
\maketitle

\abstract{
Choosing a portfolio of risky assets over time that maximizes the expected return at the same time as it minimizes portfolio risk is a classical problem in Mathematical Finance  and is referred to as the dynamic Markowitz problem (when the risk is measured by variance) or more generally, the dynamic mean-risk problem.
In most of the literature, the mean-risk problem is scalarized and it is well known that this scalarized problem does not satisfy the (scalar) Bellman's principle. Thus, the classical dynamic programming methods are not applicable. For the purpose of this paper we focus on the discrete time setup, and we will use a time consistent dynamic convex risk measure to evaluate the risk of a portfolio.

We will show that when we do not scalarize the problem, but leave it in its original form as a vector optimization problem, the upper images, whose boundary contains the efficient frontier, recurse backwards in time under very mild assumptions. 
Thus, the dynamic mean-risk problem does satisfy a Bellman's principle, but a more general one, that seems more appropriate for a vector optimization problem: a set-valued Bellman's principle.

We will present conditions under which this recursion can be exploited directly to compute a solution in the spirit of dynamic programming. Numerical examples illustrate the proposed method.
The obtained results open the door for a new branch in mathematics: dynamic multivariate programming.
\\[.2cm]
\\[.2cm]
}

\section{Introduction}

Richard Bellman introduced dynamic programming in 1954 in his seminal work~\cite{B54}. Until today, it is an essential tool that is widely used in many areas of engineering, applied mathematics, economic theory, financial economics, and natural sciences. It allows to break complicated multi-period (scalar) optimization problems into a sequence of smaller and easier sub-problems that can be solved in a recursive manner. We review the basic facts here, which make a comparison to the results of this paper easier.

Consider a time $t$ problem, for $t\in\{0,...,T-1\}$, of the following form:
Given a starting value $v_t$ of the state variable (e.g. initial wealth) at time $t$, we look for a sequence of decisions that minimizes the overall expected costs at time $t$
\begin{align}\label{dynOP}
V_t(v_t):=\min\limits_{u_t, \dots, u_{T-1}} \;\;\; & \mathbb{E}_t \, \left[ \, \sum\limits_{s=t}^{T-1} f_s(v_s, u_s, z_s)+f_T(v_T) \, \right] \\
\nonumber s.t. \;\;\; & v_{s+1} = h_s(v_s, u_s, z_s), \\
%\nonumber & x_t = x \\
\nonumber & u_s \in U_s(v_s), \\
\nonumber & z_s \in Z_s, \;\; s = t,\dots,T-1,
\end{align}
where the scalar function $f_s$ represents the costs at time $s$ of choosing the admissible control (decision) $u_s \in U_s(v_s)$, observing the random variable $z_s \in Z_s$, and obtaining the new state $v_{s+1}$ from the state equation $h_{s}$, see~\cite{B05}. One calls $V_t$ the value function of the problem and considers $v_t$, the value of the state variable, as its argument. 

The problem satisfies Bellman equation (or the Bellman's principle and is called time consistent) if the value function $V_t(v_t)$ satisfies
\begin{align}\label{Bell}
V_t(v_t) \; &= \; \min\limits_{u \in U_t(v_s)} \; \mathbb{E}_t  \left[ f_t (v_t, u, z_t) + V_{t+1} (h_t (v_t, u, z_t)) \right], %\\
%&= \; \mathbb{E}_t  \left[ f_t (x, \hat{v}_t(x), z_t) + V_{t+1} (g_t (x, \hat{v}_t(x), z_t)) \right], \\
%V_T(x) \; &= \; f_T(x), \; \text{for all } x ,
\end{align}
where we set $V_T (v_T) = f_T (v_T)$ for all $v_T$.
Then, instead of solving one complicated dynamic problem~\eqref{dynOP}, one can solve $T-1$ easier one-step problems~\eqref{Bell} backwards in time, where one uses the obtained value function $V_{t+1}$ as the input for the time $t$ problem. Equation~\eqref{Bell} has the following economic interpretation: The optimal time $t$ value $V_t$ is the sum of the optimal cash flow in the current period plus the optimal value $V_{t+1}$ in the next period. 

The term Bellman equation is used in connection with discrete-time problems. In continuous-time optimization problems, the analog is a partial differential equation called the Hamilton-Jacobi-Bellman equation. For the purpose of this paper, we will work in discrete time.

The aim of this paper is to deduce a similar Bellman equation for the mean-risk problem. The mean-risk problem has two objectives: to minimize the risk of the portfolio while maximizing the expected terminal value. Usually, a scalarization method is applied that turns the two-objective problem into a scalar one. But the obtained scalar problem does
not satisfy the Bellman equation~\eqref{Bell} and therefore turns out to be time inconsistent, see~\cite{CLWZ12, BM08}. Researchers have dealt with this problem by establishing different methods to solve this time inconsistent scalar problem.
For example, \cite{LN00} embeds the time inconsistent mean-variance problem into a one-parameter family of time consistent optimal control problems, the game theoretic interpretation of time inconsistency of~\cite{BM14} was used in~\cite{BMZ14}, a mean field approach e.g. in~\cite{AnkD11}, a dynamic change in the scalarization to turn the time inconsistent problem into a time consistent one was used in~\cite{KMZ16}, and a time-varying trade-off is combined with relaxed self-financing restrictions allowing the withdrawal of money out of the market leading to a policy dominating the precommitted one in~\cite{CLWZ12}.

We propose a completely different approach. We propose to look at the original two-objective vector optimization problem - and not at the scalarized one - and develop a Bellman equation tailored to the multi-objective nature of the problem.

The earliest results on dynamic vector-valued problems seems to be~\cite{Brown65}, where a principle of optimality for problems with values in a partially ordered multiplicative lattice are provided. \cite{Li87} solve deterministic multi-objective problems through the envelope approach, where efficient frontiers at $t+1$ for various values of the variables are expressed as parametrized curves and the non-dominated points at time $t$ are found as an envelope of the objectives of this family. \cite{Li90a} extends this to risk-neutral stochastic multi-objective problems.

The problem considered here is  a risk-averse problem and we do not assume that the efficient frontiers have a parametrized representation. However, the proposed method is interpretation-wise also in line with~\cite{Li90b}, where nonseparable scalar problems are considered that can be replaced by multi-objective problems with separable objectives.

While the previous literature provides ways to recursively solve (deterministic or risk-neutral) vector-valued dynamic problems, to the best of our knowledge, an explicit analog of~\eqref{Bell} for a vector optimization problem (VOP) was not known yet. The reason is, that it is per se not clear what the value function $V_t$ of a VOP is.
This is related to the question of what is actually meant by ``minimizing a vector function $\Gamma$''.
Classically, one tried to find all feasible points $y$ whose image $\Gamma(y)$ is efficient (i.e. non-dominated).
It did not mean to literally search for an infimum of $\Gamma$ with respect to the vector order, ``as it may not exist, and even if it does, it is not useful in practice as it refers to so-called utopia points which are typically not realizable by feasible decisions'', see~\cite{FPP15}. Thus, in the classical framework, one cannot hope to obtain a solution in the sense that the ``infimum of a vector function is attained'' and thus the ``infimum becomes a minimum" and the value of that minimum is the value $V_t$ of the problem. Thus, a value function is not defined.

This situation, however, changed drastically with the so called lattice approach to VOPs that has been introduced very recently, see~\cite{Lohne11,FPP15}. In this approach, a vector function $\Gamma$ is extended to a set-valued function $G(y)=\Gamma(y)+ \mathbb{R}^q_+$ of type ``point plus cone" and instead of a vector optimization problem w.r.t. $\Gamma$, a set-optimization problem w.r.t. $G$ is considered. This procedure is called the lattice extension of the VOP. Then, the solution concept of set-optimization, see~\cite{Lohne11,FPP15}, is applied to this particular set-optimization problem and yields a new solution concept for the original VOP.
The (lattice) infimum is now well defined, and the infimum attainment is part of this new solution concept. It turns out that the value function of a VOP in the lattice approach is nothing else than the upper image $\mathcal{P}_{t}(v_t)$ of the VOP, i.e. a set whose boundary contains the well known efficient frontier.

Now, having established a concept for the value function of a VOP, it makes  sense for the first time to try to find an analog of~\eqref{Bell} for the VOP of interest, the mean-risk problem. Since the value function of interest turned out to be a set-valued function, recent results on backward recursions and time consistency for set-valued risk measure in~\cite{FeinsteinETAL17} provided an intuition on the type of results one can expect. In contrast to the pure backward problem considered in~\cite{FeinsteinETAL17}, where a terminal random variable is given as the input, the problem here is more complicated as it is a backward-forward problem as the initial capital is provided as an input. 

One key result of this paper is to show that the upper images, i.e. the value functions $\mathcal{P}_{t}(v_t)$ recurse backwards in time, which provides a formula in total analogy to~\eqref{Bell}
\begin{align*}
%\label{Bellman2}
\begin{split}
\mathcal{P}_{t} \left( v_t \right)  = \; \inf\limits_{\substack{\trans{S_t} \psi_t = v_t,\\ \psi_t \in \Phi_t}} \Gamma_t (-\mathcal{P}_{t+1} (\trans{S_{t+1}} \psi_t)),
\end{split}
\end{align*}
or, equivalently
\begin{align}
\label{Bellmann1}
\begin{split}
\mathcal{P}_{t} \left( v_t \right) = \cl \left\lbrace  
\begin{pmatrix}
 -\mathbb{E}_t (  -x_1  ) \\
 \;\;\; \rho_t (  -x_2   )
\end{pmatrix} \right.   \Big\vert   \;\; \trans{S_t} \psi_t = v_t, \;\; \psi_t \in \Phi_t, \;\;
\left. 
\begin{pmatrix}
x_1  \\ x_2
\end{pmatrix} \in \mathcal{P}_{t+1} \left( \trans{S_{t+1}} \psi_t \right) 
\right\rbrace .
\end{split}
\end{align}

The details and notations will be introduced in the following sections. We will show that~\eqref{Bellmann1} can be rewritten as a series of one-time-step convex vector optimization problems. Solving these recursively backwards in time would solve the original dynamic mean-risk problem with upper image $\mathcal{P}_{0}(v_0)$. This is in total analogy to the scalar dynamic programming principle.

Of course, several challenges arise: How does one deal with the issue that in the backward recursion one needs to know the value function for any value $v_t$ of the state variable at time $t$? 
In some scalar and some deterministic or risk-neutral multi-objective cases this is accomplished by e.g. deriving analytical solutions as functions of $v_t$. However, in general, there is not much hope to expect analytical expressions for the solutions of vector optimization problems. Efficient algorithms exist to compute a solution and the value function of the lattice extension, but analytical expressions will be a rare exception. In this paper, the problem will be resolved by using a coherent time consistent risk measure to measure portfolio risk, this allows to scale the problem and so it suffices to solve in each node one VOP for initial value~$v_t=1$.

Two numerical examples illustrate the theoretical results. In a two-asset market the mean-risk problem is solved over $2.500$ time periods, corresponding to $10$ years of daily trading. Once the investor chooses an efficient point on the frontier that he wants to reach, the optimal trading strategy is calculated forward in time on the realized path. The second example illustrates the results in a market with multiple assets.

The efficient trading strategy moves on the efficient frontiers over time and is thus naturally related to a moving (i.e. time- and state-dependent) scalarization  that would make the scalarized problem time consistent in the scalar sense. This relates our results to the results of~\cite{KMZ16} and to the concept of \textit{time consistency in efficiency of~\cite{CLWZ12}}. The main difference is that the moving scalarization comes out implicitly as part of the solution in our approach, while in~\cite{KMZ16} it has to be found a priory, which can be done in some special cases, but was an open problem in the general case. An economic interpretation of the moving scalarization will be given in Section~\ref{sec_Scalarization}.
The method proposed in this paper recurses the efficient frontiers backwards in time, which corresponds to working with all scalarizations at the same time. However, one actually does not even need to compute the weights for this moving scalarization as one is primarily interested in the optimal trading strategy. Thus, the set-valued Bellman's principle overcomes the problematic need to explicitly compute the moving scalarization a priori, as there is no need to turn the problem into a scalar time consistent problem since the original problem can be solved already by the proposed multivariate dynamic programming principle and is thus already time consistent in the set-valued sense.
This indicates that there is a more general concept in dynamic multivariate programming that addresses some of the problems in~\cite{KMZ16}, but many open technical challenges in the general case still need to be addressed in future research. Thus, this paper can be seen as a first case study of a very general and new concept.

%%%%%%%%%%%%%%%%%%%%%%%%%%%%%%%%%%%%%
\section{The portfolio selection problem}%%%%%%%%%%%%%%%%%%%%%%%%%%%%%%%%%%%%%
%%%%%%%%%%%%%%%%%%%%%%%%%%%%%%%%%%%%%
In this section, we introduce the multi-period mean-risk problem and all basic notations and definitions.

\subsection{Preliminaries and notation}%%%%%%%%%%%%%%%%%%%%%%%%%%%%%%%%%%%%%
On a finite discrete time horizon $\mathbb{T} = \{0, 1, \dots, T\}$ consider a finite filtered probability space $( \Omega, \mathcal{F}, (\mathcal{F}_t)_{t \in \mathbb{T}}, \mathbb{P})$ with $\mathcal{F}_0$ trivial and $\mathcal{F}_T = \mathcal{F}$. Without loss of generality we assume that all nontrivial events have positive probability, i.e. $\mathbb{P} (A) > 0$ for all $A \in \mathcal{F}, A \neq \emptyset$. The set of atoms in $\mathcal{F}_t$ is denoted by $\Omega_t$.  The space of all bounded $\mathcal{F}_t$-measurable random variables is denoted by $L_t  := L_t^{\infty} ( \Omega, \mathcal{F}_t, \mathbb{P}; \mathbb{R})$. For a subset $D \subseteq \mathbb{R}^{m}$, denote the space of all bounded $\mathcal{F}_t$-measurable random vectors taking values in $D$ by $L_t(D)  := L_t^{\infty} ( \Omega, \mathcal{F}_t, \mathbb{P}; D)$. The space $L_t (\mathbb{R}^m)$ is a topological vector space; for any subset $B \subseteq L_t (\mathbb{R}^m)$ the notations $\cl B$ and $\interior B$ denote the closure and the interior, respectively. A point $\bar{x} \in B$ is called \textit{minimal} in $B$ if $\left( \bar{x} - L_t(\mathbb{R}^m_+) \backslash \{0\} \right) \cap B = \emptyset$. It is called \textit{weakly minimal} if $\left( \bar{x} - \interior L_t(\mathbb{R}^m_+) \right) \cap B = \emptyset$.

The value of a random vector $X \in L_t(\mathbb{R}^m)$ at a given atom (node) $\omega_t \in \Omega_t$ is to be understood as its value at any outcome $\omega \in \omega_t$, i.e. $X(\omega_t) := X(\omega)$. The product of two random variables is understood state-wise, $(X \cdot Y) (\omega) := X (\omega) \cdot Y(\omega)$. 
Random vectors constantly equal to $1$, resp.~$0$ are denoted by $\mathbf{1}$, resp.~$\mathbf{0}$. We do not explicitly denote their dimensions as they should be clear from the context.
For any $A \in \mathcal{F}_t$ an indicator function $I_A$ is defined as $I_A (\omega) = 1$ for $\omega \in A$ and $I_A (\omega) = 0$ otherwise.
 The conditional expectation $\mathbb{E} \left( \cdot \; \vert \;\mathcal{F}_t \right)$ is denoted by $\mathbb{E}_t \left(\cdot \right)$. 

\subsection{Market model, feasible portfolios and measurement of risk}%%%%%%%%%%%%%%%%%%%%%%%%%%%%%%%%%%%%%
A market with $d$ assets is modeled by a $d$-dimensional adapted discounted price process $(S_s)_{s = 0, \dots, T}$ on the probability space $( \Omega, \mathcal{F}, (\mathcal{F}_t)_{t \in \mathbb{T}}, \mathbb{P})$. The existence of an underlying num\'{e}raire is assumed. The distribution of the prices is assumed to be known to the investor. The probability measure $\mathbb{P}$ is not required to be the true market probability, but rather the one the investor believes to describe the market.

The investor enters the market at time $0$ with some wealth $v_0$, which is to be invested until terminal time $T$, and follows an adapted trading strategy $(\psi_s)_{s = 0, \dots, T-1 }$.
Here, $\psi_{s,i}$ denotes the number of units of an asset $i$ held in the interval between time $s$ and $s+1$. For the purposes of this work a market without transaction costs is considered. Any trading strategy the investor can follow needs to have the self-financing property, $\trans{S_s} \psi_{s} = \trans{S_s} \psi_{s-1}$ for $s = 1, \dots, T-1$. The value of the portfolio arising from a trading strategy $(\psi_s)_{s = 0, \dots, T-1 }$ is
$$v_s := \trans{S_s} \psi_{s-1},$$
for $s = 1, \dots , T$.
In the rare case when the dependency of the portfolio value from the trading strategy $\psi$ has to be made explicit, we will use the notation $v_s^{\psi}$ for $v_s$ instead.
Since the underlying probability space is finite, the portfolio value $v_s$ for any trading strategy is a bounded $\mathcal{F}_s$-measurable random variable.

Either the market authorities, or the investor herself can impose additional constraints on the positions the investor is allowed, or willing, to take. These are modeled by a sequence of constraints sets $\{ \Phi_s \}_{s = 0, \dots, T-1}$, where each $\Phi_s \subseteq L_s(\mathbb{R}^d)$ is a closed conditionally convex set. Thus, we will consider the trading restrictions
$$\psi_s \in \Phi_s \quad \text{ for } s = 0, \dots, T-1.$$
Short-selling constraints, i.e. the case $\Phi_s = L_s(\mathbb{R}^d_+)$, are studied in detail in Section~\ref{sec_implementation}. However, the derived theory works for any closed conditionally convex sets $\Phi_s$, $s = 0, \dots, T-1$.

We will use the term strategy and portfolio synonymously, and will always mean a portfolio resulting from the strategy under consideration.
An investor with wealth $v_t \in L_t$ at time $t$ will consider all possible portfolios with initial value $\trans{S_t} \psi_t = v_t$ satisfying the above conditions. We will refer to such a portfolio as feasible, and denote the set of all feasible portfolios by
\begin{align*}
\Psi_t (v_t) := \big\{ &(\psi_s)_{s = t, \dots, T-1} \; \vert \; \trans{S_t} \psi_t = v_t, \,\, \psi_t \in \Phi_t, \,\,\trans{S_s} \psi_{s-1} = \trans{S_s} \psi_{s}, \,\,\psi_s \in \Phi_s, \,\,s = t+1, \dots, T-1 \big\}.
\end{align*}

Finally, to formulate the problem, we will specify how the mean and the risk of the terminal value are measured. The mean is as usual quantified by the conditional expectation. In this paper, the risk is assessed by a dynamic time consistent convex risk measure, a concept widely used in the risk measure literature. Here we follow Sections 2 and 6 of \cite{DetlefsenETAL05} and Section 1 of \cite{Riedel04} for definitions and properties. 

\begin{definition}%[See Sections 2 and 6 of \cite{DetlefsenETAL05} and Section 1 of \cite{Riedel04}]
A \textbf{dynamic convex risk measure} is a family $(\rho_t)_{t \in \mathbb{T}}$, where every $\rho_t : L_T \rightarrow L_t$ satisfies $\rho_t (\mathbf{0}) = 0$, and for any $X, Y \in L_T$ and $\Lambda \in L_t$ the following properties hold true
\begin{itemize}
	\item \textbf{conditional translation invariance:} 
	$\rho_t (X + \Lambda) = \rho_t (X) - \Lambda$,
	\item \textbf{monotonicity:} 
	$X \leq Y \Rightarrow \rho_t (X) \geq \rho_t (Y)$,
	\item \textbf{conditional convexity:} 
	$\rho_t (\Lambda X + (1 - \Lambda)Y ) \leq \Lambda \rho_t (X) + (1 - \Lambda) \rho_t (Y)$ for $0 \leq \Lambda \leq 1$.
\end{itemize}
The dynamic convex risk measure is called \textbf{coherent} if additionally each $\rho_t$ satisfies
\begin{itemize}
	\item \textbf{conditional positive homogeneity:}  $\rho_t (\Lambda X) = \Lambda \rho_t (X)$ for $\Lambda > 0$.
\end{itemize} 
A dynamic risk measure $(\rho_t)_{t \in \mathbb{T}}$ is \textbf{time consistent}, if for any $X, Y \in L_T$ and $0 \leq t \leq T-1$ it satisfies
$\rho_{t+1} (X) = \rho_{t+1} (Y) \; \Rightarrow \; \rho_t (X) = \rho_t (Y).$
\end{definition} 

\begin{lemma}%[See Sections 4 and 6 of \cite{DetlefsenETAL05} and Section 10 in \cite{Rockafellar70}] 
\label{lemma_RM}
\begin{itemize}
\item Each element of a dynamic convex risk measure also satisfies \textbf{locality} (often also called regularity):
$\rho_t(I_A X) = I_A \rho_t(X)$ for any $A \in \mathcal{F}_t$. 
\item For every dynamic risk measure $(\rho_t)_{t \in \mathbb{T}}$ time consistency is equivalent to \textbf{recursiveness}: $\rho_t (X) = \rho_t ( - \rho_{t+1} (X))$ for any $X \in L_T$ and all $0 \leq t \leq T-1$ .
\item Every convex risk measure on a finite probability space is a continuous functional.
\item The negative conditional expectation, $-\mathbb{E}_t$, is a time consistent dynamic coherent risk measure, which is additionally linear and strictly monotone.
\end{itemize}
\end{lemma}

Throughout this paper, we will work only with time consistent dynamic convex risk measures, and will for the sake of brevity just call them risk measures. In Section~\ref{sec_implementation} we focus on time consistent dynamic coherent risk measures, and will use the term coherent risk measure then. We believe this should not lead to any misunderstanding. The assumption of time consistency of the risk measure is reasonable as it assures that the investor's risk assessment does not contradict itself over time.

Summarizing, we formulate the assumptions posed on the market and the investor, whose point of view is adopted throughout.

\begin{assumption}  $\;$ 
\label{Assumption1}
\begin{enumerate}
\item The investor's perception of the market is represented by the adapted discounted price process $(S_s)_{s = 0, \dots, T}$ on a finite filtered probability space $( \Omega, \mathcal{F}, (\mathcal{F}_t)_{t \in \mathbb{T}}, \mathbb{P})$. The distributions of the prices are known.
\item The investor with wealth $v_t \in L_t$ at time $t$ considers only portfolios $(\psi_s)_{s = t, \dots, T-1 }$ which
have initial value $v_t$, are self-financing, and satisfy $\psi_s \in \Phi_s \text{ for } s = t, \dots, T-1$ for given closed conditionally convex sets $\Phi_s \subseteq L_s(\mathbb{R}^d)$ modeling trading constraints. These three conditions form the set of feasible portfolios $\Psi_t (v_t)$.
\item The investor enters the market with wealth $v_0$ at time $0$, which is to be invested there until terminal time $T$. It is assumed that $\Psi_0 (v_0) \neq \emptyset$.
\item The investor evaluates portfolios by the mean and the risk of their terminal values $v_T$, where
\begin{enumerate}
\item the mean is assessed by the conditional expected value $\mathbb{E}_t$,
\item the risk is quantified by a time consistent dynamic convex risk measure $(\rho_t)_{t \in \mathbb{T}}$. 
\end{enumerate}
\end{enumerate}
\end{assumption}

The non-emptiness of the set of feasible portfolios is only assumed at initial time~$0$. In Lemma~\ref{Lemma_recursivePsi} below we will show that this implies the non-emptiness of the set of feasible portfolios $\Psi_t (v_t)$ for all relevant investments $v_t$ that can be reached from wealth $v_0$.

\subsection{Efficient portfolios}%%%%%%%%%%%%%%%%%%%%%%%%%%%%%%%%%%%%%
A rational investor will only choose among non-dominated portfolios, so called efficient portfolios. This concept in the setting of Assumption~\ref{Assumption1} will now be made precise together with the broader concept of weak efficiency covering also portfolios which are not strictly dominated.  
Since the investor can make decisions dynamically, the concept of efficiency is defined for every time point.

\begin{definition}
\label{def_efficient}
Under Assumption~\ref{Assumption1}, a feasible portfolio $(\psi_s)_{s = t, \dots, T-1} \in \Psi_t (v_t)$ is called \textbf{ time $\mathbf{t}$ efficient for initial wealth $\mathbf{v_t}$} if, and only if, there exists no other feasible portfolio $(\phi_s)_{s = t, \dots, T-1} \in \Psi_t (v_t)$, such that
\begin{align}
\label{def_eff_ineq}
\begin{split}
\mathbb{E}_t \left( v_T^{\phi} \right) &\geq \mathbb{E}_t \left( v_T^{\psi} \right), \\
\rho_t \left( v_T^{\phi} \right) &\leq \rho_t \left( v_T^{\psi} \right),
\end{split}
\end{align}
where at least one of the above inequalities is a strict inequality $\mathbb{P}$-a.s. The set of all such portfolios is called the time $t$ efficient frontier for initial wealth $v_t$.

A feasible portfolio $(\psi_s)_{s = t, \dots, T-1} \in \Psi_t (v_t)$ is called \textbf{ time $\mathbf{t}$ weakly efficient for initial wealth $\mathbf{v_t}$} if both inequalities in \eqref{def_eff_ineq} are strict for all $\omega_t\in\Omega_t$. 
\end{definition}

Efficiency is also called Pareto optimality and weak efficiency weak Pareto optimality. The term efficient frontier will also be used for the set of all objective values of efficient portfolios.

\begin{remark}
Note that the strict inequalities in the definitions of efficiency and weak efficiency differ as the first one is understood in the $\mathbb{P}$-a.s. sense (i.e. $X<Y$ $\mathbb{P}$-a.s. iff $X\leq Y$ and $\mathbb{P}(X<Y)>0$) and the second one is omega-wise. The mathematical intuition behind this will become clear at the end of this subsection, when we relate (weak) efficiency to the order relation. For an economic interpretation note that for an efficient portfolio there cannot exist another portfolio that is not worse, but better in at least one component in at least one node. For a weakly efficient portfolio there should not be a portfolio that is better in all components in all states.
\end{remark}

One may immediately notice  in the above definition the dependency of efficiency on the wealth $v_t$. This is necessary, since in general it is not possible to derive an explicit relation between the efficient frontiers for different wealths. The situation simplifies when the risk measure is coherent. Then, the efficient frontiers scale, which is discussed in Subsection~\ref{sec_coherent}. The viewpoint of a single node $\omega_t \in \Omega_t$ for the definitions of (weak) efficiency is discussed in the e-companion in Section~\ref{EC_node}.

To assign to each portfolio its mean-risk profile, we define
a vector-valued function $\Gamma_t: L_T (\mathbb{R}^2 ) \to L_t (\mathbb{R}^2 )$, which applies the negative conditional expectation and the risk measure component-wise to a random vector, that is 
\begin{align*}
\Gamma_t (X) := 
\begin{pmatrix}
- \mathbb{E}_t \left( X_1 \right) \\ 
\;\;\; \rho_t \left( X_2 \right)
\end{pmatrix}.
\end{align*}
For any feasible portfolio $\psi \in \Psi_t (v_t)$, the investor is at time $t$ interested in the value 
$\Gamma_t (V_T (\psi))$, where
\begin{align}
\label{terminalwealth}
V_T (\psi) 
= \begin{pmatrix}
v_T \\ v_T
\end{pmatrix} = \begin{pmatrix}
\trans{S_T} \psi_{T-1} \\ \trans{S_T} \psi_{T-1}
\end{pmatrix}
%= \trans{\left( v_T, v_T \right)} = \trans{\left( \trans{S_T} \psi_{T-1} , \trans{S_T} \psi_{T-1} \right)}
\end{align} 
is a two dimensional vector of the terminal wealth. The reason for defining $\Gamma_t$ as a function of a random vector $V_T(\psi)$, rather than a random variable $v_T$ is a subsequent recursive form, which will appear in Section~\ref{sec_Bellman}.
The function $\Gamma_t(V_T(\cdot))$ is in a natural way connected to the definition of efficiency - the reader can easily convince himself that the ordering corresponding to Definition~\ref{def_efficient} is $\leq_{L_t (\mathbb{R}^2_+ )}$, and that the condition for  time $t$ efficiency is equivalent to
\begin{align}
\label{def_eff_ineq_gamma}
\nexists \phi \in \Psi_t(v_t): \;
\Gamma_t (V_T (\phi)) \leq_{L_t (\mathbb{R}^2_+ )} \Gamma_t (V_T (\psi)) \text{ and } \Gamma_t (V_T (\phi)) \neq \Gamma_t (V_T (\psi)),
\end{align}
and the condition for time $t$ weak efficiency corresponds to
\begin{align}
\label{def_eff_ineq_gamma2}
\nexists \phi \in \Psi_t(v_t): \;
\Gamma_t (V_T (\phi)) \in \left( \Gamma_t (V_T (\psi)) - \interior L_t(\mathbb{R}^2_+)  \right) .
\end{align} 
Note, that we will often use the short hand notation $\psi := (\psi_s)_{s = t, \dots, T-1 }$ for the trading strategy. We believe the initial time $t$ is clear from the context, and this ambiguity is out-weighted by the increased readability of the formulas. 
Since $\leq_{L_t (\mathbb{R}^2_+ )}$ corresponds to the natural element-wise ordering in $L_t(\mathbb{R}^2)$, it will usually be denoted by $\leq$. The ordering cone $L_t (\mathbb{R}^2_+ )$ will only be stressed in the context of the optimization problem.

\subsection{Mean-risk as a vector optimization problem}%%%%%%%%%%%%%%%%%%%%%%%%%%%%%%%%%%%%%

The investor naturally prefers the efficient portfolios, therefore wishes to maximize the mean and minimize the risk - or simply to minimize the vector-valued function $\Gamma_t$ of the terminal wealth. Our approach to portfolio selection is to formulate, and to study, the mean-risk as a vector optimization problem (VOP). Within this framework the mean-risk problem of the investor with wealth $v_t \in L_t$ at time $t$ is
\begin{align}
	\label{Probt}
	\tag{$D_t (v_t)$}
	\begin{split}
	\min\limits_{(\psi_s)_{s = t, \dots, T-1}} \;\; & 
	\begin{pmatrix}
	 - \mathbb{E}_t (  v_T  ) \\
	 \;\;\; \rho_t (  v_T  )
	\end{pmatrix} \text{ w.r.t. } \leq_{L_t (\mathbb{R}^2_+)} \\
	\text{s.t. } \;\;\;\;\;\;\;\;\;\; & \;\; \trans{S_s} \psi_s = v_s, \\		
	& \;\; v_{s+1} = \trans{S_{s+1}} \psi_s, \\
	& \;\; \psi_s \in \Phi_s, \\
	& \;\; s = t, \dots, T-1.
	\end{split}
\end{align}
Using the notation of the bi-objective function $\Gamma_t$ and the terminal wealth $V_T$, as well as the set of feasible portfolios $\Psi_t (v_t)$ as defined in Assumption~\ref{Assumption1}, problem~\ref{Probt} can be written as
\begin{align*}
\min \;\;\; &\Gamma_t (V_T (\psi)) \text{ w.r.t. } \leq_{L_t (\mathbb{R}^2_+)} \\
\text{s.t. } \; & \psi \in \Psi_t (v_t).
\end{align*}

\begin{remark}
\label{remark_VOP}
The set $L_t (\mathbb{R}^2)$ is a vector space and its subset $L_t (\mathbb{R}^2_+)$ is a pointed convex cone, which is additionally closed and solid. Thus, the pair $(L_t (\mathbb{R}^2), \leq_{L_t (\mathbb{R}^2_+)})$ is a partially ordered vector space, and thus a suitable image space for a vector optimization problem. The set  $\Psi_t (v_t)$ is closed, as it is determined via equalities and inclusion in closed sets.
Therefore as long as the feasible set $\Psi_t (v_t)$ is non-empty, problem~\ref{Probt} is a VOP, as defined in \cite{Lohne11}.

Since we are working on a finite probability space, the sets $L_t (\mathbb{R}^2)$, resp. $L_t (\mathbb{R}^d)$, are finite dimensional, and therefore isomorphic to the Euclidean space for some appropriate dimension. Consequently the mean-risk problem~\ref{Probt} can be seen as a VOP with image space $\mathbb{R}^q$ and variable space $\mathbb{R}^m$ with appropriate dimensions.
\end{remark}

The mean-risk problem is formulated for every time point $t$ and for any wealth $v_t \in L_t$. Together, these problems compose a family of mean-risk problems
$$\mathcal{D} = \big\{ D_t (v_t) \; \vert \; t \in \{0, \dots, T-1 \}, v_t \text{ is } \mathcal{F}_t\text{-measurable} \big\},$$
which will be the central object of this work. This family of problems can be interpreted in terms of dynamic programming: we study a dynamic system, a portfolio, which is at each time point $t$ described by its value, the state variable. The decision maker, in this case the investor, influences the portfolio at each time point by her choice of positions in the individual assets (the trading strategy), which is the control variable. Afterwards the market impacts the portfolio by a random change in the stock prices, which are the random shocks to the system. Our problem differs from standard dynamic programming only by considering two objectives simultaneously.

Since each problem~\ref{Probt} is a VOP, all of the concepts from vector optimization are relevant for it. The following four notions will be used in the subsequent sections. The image of the feasible set of problem~\ref{Probt} is denoted by 
\begin{align*}
\Gamma_t \left( \Psi_t (v_t) \right) := \big\{ \Gamma_t (V_T (\psi)) \; \vert \; \psi \in \Psi_t (v_t) \big\}.
\end{align*}
The upper image of~\ref{Probt} will be denoted by
\begin{align*}
\mathcal{P}_t (v_t) := \cl \left( \Gamma_t \left( \Psi_t (v_t) \right) + L_t (\mathbb{R}^2_+) \right).
\end{align*}
A feasible portfolio $\psi \in \Psi_t (v_t)$ is a minimizer of problem~\ref{Probt} if
\begin{align}
\label{def_minimizer}
\left( \Gamma_t (V_T (\psi)) - L_t(\mathbb{R}^2_+) \backslash \{\mathbf{0}\} \right) \cap \Gamma_t \left( \Psi_t (v_t) \right) = \emptyset,
\end{align}
and it is a weak minimizer if
\begin{align}
\label{def_weak_minimizer}
\left( \Gamma_t (V_T (\psi)) - \interior L_t(\mathbb{R}^2_+)  \right) \cap \Gamma_t \left( \Psi_t (v_t) \right) = \emptyset.
\end{align} 
The following lemma points out the connection between minimizers and efficient portfolios.

\begin{lemma}
\label{lemma_minimizer}
A feasible trading strategy $(\psi_s)_{s = t, \dots, T-1}$ is a (weakly) efficient portfolio at time $t$ for initial wealth $v_t$ if, and only if, it is a (weak) minimizer of problem~\ref{Probt}.
\end{lemma}

%\proof{Proof.} 
\begin{proof}
Relations~\eqref{def_eff_ineq_gamma},~\eqref{def_eff_ineq_gamma2} show that~\eqref{def_minimizer},~\eqref{def_weak_minimizer} are equivalent to the Definition~\ref{def_efficient}.
\end{proof}
%\Halmos \endproof

Consequently, the efficient frontier, as well as the weakly efficient frontier are contained in the boundary of the upper image. Associating an efficient portfolio to every minimal point of the upper image is possible when a compact feasible set is considered, this will be discussed in Lemma~\ref{lemma_bounded_efficient}. Similarly to efficiency, the optimization problem can be formulated in a node-wise fashion - this is discussed in Section~\ref{EC_node}.
As in the scalar case, convexity is a desirable property for an optimization problem. 

\begin{restatable}{lemma}{lemmaconvexPt}
\label{lemma_convexPt}
Each mean-risk problem~\ref{Probt} is a convex vector optimization problem. 
The feasible set $\Psi_t(v_t)$ and the objective function $\Gamma_t(V_T(\cdot))$ are conditionally convex. Furthermore, the objective function has the locality property.
\end{restatable}
 
The proof will be given in the Appendix~\ref{EC_proofs}.

%%%%%%%%%%%%%%%%%%%%%%%%%%%%%%%%%%%%%
\section{Time consistency}%%%%%%%%%%%%%%%%%%%%%%%%%%%%%%%%%%%%%
\label{sec_TimeConsistency}
%%%%%%%%%%%%%%%%%%%%%%%%%%%%%%%%%%%%%

Time consistency is a central issue in the fields of optimal control and risk averse dynamic programming, however, there 
are slightly varying definitions that are used for this concept.
In the context of efficient portfolios we decided to follow the approach used in \cite{RudloffETAL14}: \textit{a policy is time consistent if, and only if, the future planned decisions are actually going to be implemented.} 
Or formulated differently - understanding that one only implements what is optimal - if the optimal policy is still optimal at all later time points w.r.t. the objectives at these times.
In the portfolio selection setting, the investor wishes to choose a (weakly) efficient portfolio every time she makes a decision. It is reasonable then to assume that she will not implement any trading strategy which is not, at the moment, at least weakly efficient. This motivates the following definition.

\begin{definition}
\label{def_time_consistency}
The family of mean-risk problems $\mathcal{D}$ is called \textbf{time consistent} w.r.t. weak minimizers if and only if for each time $t=0, \dots, T-1$, $(\psi_s)_{s = t, \dots, T-1}$, being a weak minimizer of~\ref{Probt} implies $(\psi_s)_{s = t+1, \dots, T-1}$ being a weak minimizer of $D_{t+1} (\trans{S_{t+1}} \psi_t)$.
\end{definition}

Definition~\ref{def_time_consistency} is closely related to \textit{time consistency in efficiency} defined in~\cite{CLWZ12}. They do not coincide completely, as in our setting one might need to place zero weight on the risk component instead of the expectation (see Lemma~\ref{lemma_tc_one_step} and~\ref{lemma_exists_w}).

By Lemma~\ref{lemma_minimizer} our definition of time consistency applies the priciple of optimality to the  weakly efficient frontiers. Theorem~\ref{thm_time_consistency2} shows that the mean-risk problems satisfy this notion of time consistency, where Lemma~\ref{Lemma_recursivePsi} provides the recursiveness of the feasible set.

\begin{restatable}{lemma}{LemmarecursivePsi}
\label{Lemma_recursivePsi}
A trading strategy is feasible, that is $(\psi_s)_{s = t, \dots, T-1} \in \Psi_t (v_t)$ if, and only if, $\trans{S_t} \psi_t = v_t, \psi_t \in \Phi_t$, and $(\psi_s)_{s = t+1, \dots, T-1} \in \Psi_{t+1} (\trans{S_{t+1}} \psi_t)$.
\end{restatable} 
The proof will be given in the Appendix~\ref{EC_proofs}.

\begin{theorem}
\label{thm_time_consistency2}
Under Assumption~\ref{Assumption1}, the family of mean-risk problems $\mathcal{D}$ is time consistent w.r.t. weak minimizers (see Definition~\ref{def_time_consistency}).
\end{theorem}

\begin{proof}
%\proof{Proof.} 
Consider some time point $t$, an investment $v_t$ and any weak minimizer $(\psi_s)_{s = t, \dots, T-1}$ of problem~\ref{Probt}. By Lemma~\ref{Lemma_recursivePsi} the truncated trading strategy is feasible for problem $D_{t+1} (\trans{S_{t+1}} \psi_t)$, assume it  is not a weak minimizer. Then there exists a feasible trading strategy $(\phi_s)_{s \in \{t+1, \dots, T-1\}} \in \Psi_{t+1} (\trans{S_{t+1}} \psi_t)$, such that
\begin{align}
\label{eq_2}
\begin{split}
-\mathbb{E}_{t+1} \left( v_T^{\phi} \right) &< -\mathbb{E}_{t+1} \left( v_T^{\psi} \right),\\
 \rho_{t+1} \left(v_T^{\phi} \right) &< \rho_{t+1} \left(v_T^{\psi} \right),
 \end{split}
\end{align}
for all $\omega_{t+1}\in\Omega_{t+1}$.
By defining additionally $\phi_t := \psi_t$, a feasible $(\phi_s)_{s = t, \dots, T-1} \in \Psi_t (v_t)$ is obtained. Let us look at the values of the objectives for this portfolio. The tower property and the strict monotonicity of the expectation combined with~\eqref{eq_2} yields 
\begin{align}
\label{eq_3}
-\mathbb{E}_{t} \left( v_T^{\phi} \right) = \mathbb{E}_t \left(-\mathbb{E}_{t+1}  \left( v_T^{\phi} \right) \right) < \mathbb{E}_t \left( -\mathbb{E}_{t+1} \left( v_T^{\psi} \right) \right) = -\mathbb{E}_{t} \left( v_T^{\psi} \right).
\end{align}
Define $\epsilon := \min\limits_{\omega \in \Omega}\left( \rho_{t+1} (v_T^{\psi}) - \rho_{t+1} (v_T^{\phi}) \right) > 0$, then $\rho_{t+1} (v_T^{\phi}) \leq \rho_{t+1} (v_T^{\psi}) - \epsilon \mathbf{1}$. Combining this inequality, the monotonicity, the translation invariance and the recursiveness of the risk measure yields
\begin{align}
\label{eq_4}
\begin{split}
\rho_{t} (v_T^{\phi}) &= \rho_t (-\rho_{t+1} (v_T^{\phi})) \leq \rho_t (-\rho_{t+1} (v_T^{\psi}) + \epsilon \mathbf{1}) \\
 &= \rho_t (-\rho_{t+1} (v_T^{\psi})) - \epsilon \mathbf{1} = \rho_t (v_T^{\psi}) - \epsilon \mathbf{1} < \rho_t (v_T^{\psi}).
 \end{split}
\end{align}
Together,~\eqref{eq_3} and~\eqref{eq_4} contradict to $(\psi_s)_{s =t, \dots, T-1}$ being a weak minimizer. Therefore, the assumption cannot hold and $\mathcal{D}$ must be time consistent w.r.t.  weak minimizers. 
%\Halmos \endproof
\end{proof}

Notice that throughout the proof of Theorem~\ref{thm_time_consistency2} only the properties of recursiveness, monotonicity and translation invariance of the risk measure were used, but convexity was not needed. Indeed, the convexity of the risk measure was only necessary for proving the convexity of the vector optimization problem~\ref{Probt} in Lemma~\ref{lemma_convexPt}. Let us shortly consider the time consistent dynamic version of the Value at Risk, see \cite{CheriditoETAL09} for details, and let us denote it by VaR. It lacks convexity, but has otherwise all the properties assumed. Naturally, if the risk is measured by VaR, the mean-risk problems are not convex, but the proof of Theorem~\ref{thm_time_consistency2} works also in that case. Thus, the mean-VaR problem is time consistent w.r.t. weak minimizers.

The trouble with weak efficiency is that a weakly efficient portfolio is not necessarily weakly efficient in every node, see the discussion in Section~\ref{EC_node}. %Remark~\ref{remark_node_wise1}. 
Since the investor is ultimately interested in the realized path (nodes), this makes weak efficiency seem rather insufficient. One could in total analogy to Definition~\ref{def_time_consistency} define \textit{time consistency w.r.t. minimizers}, which would be the property desired by the investor. It also corresponds to the principle of optimality used in the risk-neutral and deterministic case e.g. in~\cite{Li90a} and \cite{Li87}. Unfortunately this property does not hold for the mean-risk problem in general. A sufficient condition guaranteeing it is strict monotonicity of the risk measure, which is satisfied for example for the entropic risk measure, but is in general a rather strong assumption as e.g. tail-based risk measures lack it. In Section~\ref{sec_coherent} a property stronger than time consistency w.r.t. weak minimizers, but weaker than time consistency w.r.t. minimizers will be proven under  additional assumptions (coherent risk measure and short-selling constraints). Then, for any chosen minimal mean-risk profile at time $t=0$, there exists a trading strategy, which stays efficient at all times. But even in the general setting we can obtain a result that guarantees at least weak efficiency in every node. In Lemma~\ref{lemma_tc_weaknode} in Section~\ref{EC_node} on the node-wise approach it is shown that an efficient portfolio is at each subsequent time at least weakly efficient in every node.

%%%%%%%%%%%%%%%%%%%%%%%%%%%%%%%%%%%%%%%%%%%%%%%%%%%%%%%%%%%%%%%
\section{Scalarization}%%%%%%%%%%%%%%%%%%%%%%%%%%%%
%%%%%%%%%%%%%%%%%%%%%%%%%%%%%%%%%%%%%%%%%%%%%%%%%%%%%%%%%%%%%%
\label{sec_Scalarization}
In this section we relate minimizers of the mean-risk problem to weighted sum scalarizations of the problem.
This leads to a connection between efficient portfolios and the investor's risk aversion, and allows to relate our results to existing result in the literature on mean-variance problems. 

\begin{restatable}{lemma}{movingScal}
\label{moving_Scal}
Under Assumption~\ref{Assumption1}, to every portfolio $(\psi_s)_{s = 0, \dots, T-1}$ efficient at time $0$ for wealth $v_0$ corresponds a sequence of weights $(w_s)_{s = 0, \dots, T-1}$ such that at every time $t$
\begin{itemize}
\item $w_t \in L_t(\mathbb{R}^2_+ \backslash \{0\})$ and 
\item the portfolio $(\psi_s)_{s = t, \dots, T-1}$ is an optimal solution of a weighted sum scalarization of problem $D_t (\trans{S_t} \psi_{t-1})$ with weight $w_t$.
\end{itemize}
\end{restatable}  
The proof will be given in the Appendix~\ref{EC_proofs}.

Since the weight $w_t(\omega_t)$ can be normalized, it can be interpreted as a risk aversion. The sequence $(w_s)_{s = 0, \dots, T-1}$ then represents a \textit{time-varying state-dependent risk aversion}. The given portfolio is then optimal at every time for an investor with this risk aversion solving a scalar mean-risk problem.

We can relate this to the results of \cite{KMZ16}, where it is shown that a scalar mean-variance problem can be made time consistent by a correct choice of the time-varying state-dependent risk aversions. Here, we obtained a sequence of time-varying state-dependent weights (or risk aversion), which make a portfolio efficient over time for the corresponding scalarized problem.

Note that this time-varying state-dependent risk aversion, which exists for each efficient portfolio and  makes the scalarized problem time consistent in the classical (scalar) sense, has also an economic interpretation. At time $t=0$ the investor makes a choice about the expected return and the risk she is willing to take. As time passes the market moves (either in her favor or not) and thus has an impact on the overall expectation and risk. If e.g. the market is moving in the favor for the investor she is able to 'cash in' already part of her desired expected return and can now be more relaxed about it and still be consistent with her initial choice. The same holds true for the risk she chose. Thus, an investor makes a choice about her risk aversion only at time $t=0$ and then the movement of the market determines her residual risk aversion at any time $t>0$ that is consistent with her initial choice and the part of the expected return and the risk that is already realized up to time $t$.
This interpretation is in strong contrast to the classical view in the literature, where the investor chooses at each time her risk aversion (typically the same risk aversion). From our point of view it is clear why this classical approach leads to a time inconsistent problem in the scalar sense as her new decision typically contradicts her decision made at earlier time points. 

Note that also for the classical mean-variance problem there are economic reasons why the risk aversion should not be constant over time, see \cite{BM14,BMZ14}, where a non-constant but predetermined risk aversion is chosen which still leads to a time-inconsistent problem. In contrast, \cite{KMZ16} determine the moving scalarization that turns the problem into a time consistent one which thus relates directly to out approach.

The advantage of our approach is, that one does not have to (and often can not) calculate this time-varying state-dependent risk aversion a priori, it can rather be seen as an output of our approach, as it implicitly, by Lemma~\ref{moving_Scal}, corresponds to the optimal trading strategy.

%%%%%%%%%%%%%%%%%%%%%%%%%%%%%%%%%%%%%%%%%%%%%%%%%%%%%%%%%%%%%%%
\section{Recursiveness and a set-valued Bellman's principle}%%%%%%%%%%%%%%%%%%%%%%%%%%%%
%%%%%%%%%%%%%%%%%%%%%%%%%%%%%%%%%%%%%%%%%%%%%%%%%%%%%%%%%%%%%%
\label{sec_Bellman}
In scalar dynamic programming time consistency is closely related to the famous Bellman's principle, which provides a recursive relation for the so-called value function of the problem. In the scalar setting the value function simply maps the state (in our case the wealth) to the infimum of the values the objective can attain. In this work a dynamic problem is studied, which has already been demonstrated to be time consistent. It is then natural to wonder whether the Bellman's principle holds for the mean-risk problem. However, to answer this a different question arises - how would a Bellman's principle look like for a mean-risk problem with vector-valued objective? And what would be the value function for this VOP? In this section we will answer these questions. 

\subsection{The value function of the mean-risk vector optimization problem}%%%%%%%%%
Naturally, the value function should be an infimum as in the scalar case. However, the infimum in the classical sense of the vector ordering has some well-known drawbacks - it is often an "utopia point", it provides us with little information about the problem, and for some partially ordered vector spaces it might not even exist. These were also the reasons why recently the set optimization approach was used for defining a new solution concept for VOPs based, in total analogy to the scalar case, on infimum attainment and minimality, see e.g. \cite{Lohne11}. 
 This suggest a different candidate for the value function is needed. It turns out that the infimum appearing in the set optimization approach to VOP, see \cite{FPP15}, provides a perfect candidate for a value function, and it has already been introduced here - the upper image.

Why is the upper image an infimum?  In the set optimization approach to VOP one considers a "setified" objective function
\begin{align*}
G_t (V_T(\psi)) := \Gamma_t (V_T(\psi)) + L_t (\mathbb{R}^2_+).
\end{align*}
This is a set-valued function mapping into space of closed upper sets $\mathbb{F} := \mathbb{F} \left( L_t (\mathbb{R}^2), L_t (\mathbb{R}^2_+) \right) = \big\{ A \subseteq L_t (\mathbb{R}^2) \; \vert \; \cl (A + L_t (\mathbb{R}^2_+)) = A \big\}$. The space $\mathbb{F}$ is a conlinear space (see \cite{FPP15}) with a partial ordering~$\supseteq$. The pair $(\mathbb{F}, \supseteq)$ is a complete lattice, and the infimum of a subset $\mathbb{A} \subseteq \mathbb{F}$ is given by
\begin{align*}
\inf\limits_{(\mathbb{F}, \supseteq)} \mathbb{A} = \cl \bigcup\limits_{A \in \mathbb{A}} A.
\end{align*}
Details on the theory of set optimization can be found in \cite{FPP15}. By replacing the vector-valued objective function $\Gamma_t$ in the mean-risk problem~\ref{Probt} by the set-valued objective function $G_t$, a set-valued mean-risk problem
\begin{align*}
\min \;\;\; &G_t (V_T(\psi)) \text{ w.r.t. } \supseteq \\
\text{s.t. } \; & \psi \in \Psi_t (v_t)
\end{align*}
is obtained. This set-valued problem is closely related to the original vector-valued problem since both have the same feasible points and the same minimizers, see Chapter~7.1 in \cite{FPP15}. 
The infimum of the mean-risk problem in $(\mathbb{F}, \supseteq)$ turns out to be the upper image as
\begin{align*}
\inf\limits_{\psi \in \Psi_t (v_t)} \; G_t (V_T(\psi)) &= \cl  \bigcup\limits_{\psi \in \Psi_t (v_t)} \left( \Gamma_t (V_T(\psi)) + L_t (\mathbb{R}^2_+) \right) \\
&= \cl \left( \Gamma_t \left( \Psi_t (v_t) \right) + L_t (\mathbb{R}^2_+) \right) = \mathcal{P}_t (v_t).
\end{align*}

\subsection{A set-valued Bellman's principle}%%%%%%%%%%%%%%
\label{sec_Bellmanp}

Since the upper image corresponds to the infimum of the VOP in the set-valued sense, it is a suitable candidate for a value function. The Bellman's principle for the mean-risk problem should then be a recursive relation expressing the upper image of the problem~\ref{Probt} via the upper images of the mean-risk problems at time $t+1$.
The following theorem is the main result of this section and will provide the recursiveness of the value function, i.e. the upper image, and thus establishes a Bellman equation for the mean-risk problem.

\begin{theorem}
\label{theorem_Bellman}
The upper images of the mean-risk problems~\ref{Probt} have a recursive form 
\begin{align}
\label{Bellman}
\begin{split}
\mathcal{P}_{t} \left( v_t \right) = \cl \left\lbrace  
\begin{pmatrix}
 -\mathbb{E}_t (  -x_1  ) \\
 \;\;\; \rho_t (  -x_2   )
\end{pmatrix} \right.   \Big\vert   \;\; \trans{S_t} \psi_t = v_t, \;\; \psi_t \in \Phi_t, \;\;
\left. 
\begin{pmatrix}
x_1  \\ x_2
\end{pmatrix} \in \mathcal{P}_{t+1} \left( \trans{S_{t+1}} \psi_t \right) 
\right\rbrace .
\end{split}
\end{align}
\end{theorem}

%\proof{Proof.} 
The proof will be given in the Appendix, Section~\ref{sec proof T1}.
%\Halmos \endproof

Related to representation~\eqref{Bellman} is a one-time-step optimization problem
\begin{align}
\label{ProbRRt}
\tag{$\tilde{D}_t (v_t)$}
\begin{split}
\min\limits_{\psi_t, (x_1, x_2)} \;\; & 
\begin{pmatrix}
 -\mathbb{E}_t (  -x_1  ) \\
 \;\;\; \rho_t (  -x_2   )
\end{pmatrix} \text{ w.r.t. } \leq_{L_t (\mathbb{R}^2_+)}  \\
\text{s.t. } \;\, & \;\; \trans{S_t} \psi_t = v_t, \\
& \;\; \psi_t \in \Phi_t, \\
& \;\; 
\begin{pmatrix}
x_1  \\ x_2
\end{pmatrix} \in \mathcal{P}_{t+1} \left( \trans{S_{t+1}} \psi_t \right).
\end{split}
\end{align}
For the problem to be well-defined for all times, including the pre-terminal time $T-1$, we define the set $\mathcal{P}_T (v_T)$, which depends on the $\mathcal{F}_T$-measurable input $v_T,$ as
\begin{align}
\label{upperT}
%\label{setTT}
\mathcal{P}_T (v_T) := 
%\Gamma_{T} \left( \Psi_{T} (v_T)  \right) + L_T(\mathbb{R}^2_+)  = 
\left\lbrace \begin{pmatrix}
-v_T \\ -v_T
\end{pmatrix} \right\rbrace + L_T(\mathbb{R}^2_+) .
\end{align}
We will prove the following two properties of this one-time-step optimization problem~\ref{ProbRRt}, which will justify why relation~\eqref{Bellman} can be called a  Bellman equation for the mean-risk problem.

\begin{lemma}
\label{thm_identical_upper_images}
The upper image $\tilde{\mathcal{P}}_t \left( v_t \right)$ of problem~\ref{ProbRRt} coincides with the upper image of the original mean-risk problem~\ref{Probt}, that is
\begin{align}
\label{eq_upper_images}
\tilde{\mathcal{P}}_t \left( v_t \right)=\mathcal{P}_{t} \left( v_t\right).
\end{align}
\end{lemma}
%\proof{Proof.} 
The proof will be given in the Appendix, Section~\ref{sec proof L1}.
%\Halmos \endproof

\begin{lemma}
\label{lemma_recursive_CVOP}
Problem~\ref{ProbRRt} is a convex vector optimization problem.
\end{lemma}

%\proof{Proof.} 
The proof will be given in the Appendix, Section~\ref{sec proof L2}.
%\Halmos \endproof

These two properties ensure, that a series of one-time-step convex vector optimization problems~\ref{ProbRRt} can be solved backwards in time in order to solve the original dynamic mean-risk problem~\ref{Probt} with upper image $\mathcal{P}_{t} \left( v_t\right)$. This is in total analogy to the scalar dynamic programming principle, where a complicated dynamic problem can be chopped into smaller one-time-step problems that are then solved backwards in time. The only difference here is, that instead of a scalar optimization problem, a convex VOP is solved at each point in time. Algorithms to solve convex VOPs like \cite{LohneETAL14} (or \cite{{RudloffETAL17,HamelLohneRudloff14,Lohne11,RusVanderbei03}} in the linear case) can be used, which compute a solution to the VOP in the sense of \cite{Lohne11,LohneETAL14}, but they also compute the upper image, which will then be used as an input for the constraints of the optimization problem at the next earlier time point.

Lastly, we provide an interpretation of~\eqref{Bellman} as a recursive infimum which is  in total analogy to the scalar case~\eqref{Bell}. By Theorem~\ref{theorem_Bellman} and Lemma~\ref{thm_identical_upper_images}, relation~\eqref{Bellman} can be rewritten as
\begin{align}
\label{BellmanB}
\begin{split}
\mathcal{P}_{t} \left( v_t \right) %&= 
%\cl \left\lbrace  
%\begin{pmatrix}
 %-\mathbb{E}_t ( \; -x_1 \; ) \\
 %\rho_t ( \; -x_2 \;  )
%\end{pmatrix} \right.   \Big\vert   \;\; \trans{S_t} \psi_t = v_t, \psi_t \in \Phi_t, \;\;
%\left. 
%\begin{pmatrix}
%x_1  \\ x_2
%\end{pmatrix} \in \mathcal{P}_{t+1} \left( \trans{S_{t+1}} \psi_t \right) 
%\right\rbrace  
%\\
%& \; \\
%&
= \cl \bigcup_{\substack{ \trans{S_t} \psi_t = v_t, \psi_t \in \Phi_t, \\ x  \in \mathcal{P}_{t+1} \left( \trans{S_{t+1}} \psi_t \right) }} \Gamma_t ( -x ). 
%\\
%	& \; \\
%	&
%	\;=\; \inf_{\substack{ \trans{S_t} \psi_t = v_t, \psi_t \in \Phi_t, \\ x  \in \mathcal{P}_{t+1} \left( \trans{S_{t+1}} \psi_t \right) }} G_t ( -x ).
%\; = \; \inf\limits_{\substack{\trans{S_t} \psi_t = v_t,\\ \psi_t \in \Phi_t}} \Gamma_t (-\mathcal{P}_{t+1} (\trans{S_{t+1}} \psi_t)),
\end{split}
\end{align}
Now define
$$\Gamma_t (-\mathcal{P}_{t+1} (\trans{S_{t+1}} \psi_t)) := \cl \bigcup\limits_{x \in  \mathcal{P}_{t+1} (\trans{S_{t+1}} \psi_t)} \Gamma_t (-x) = \inf\limits_{x \in  \mathcal{P}_{t+1} (\trans{S_{t+1}} \psi_t)} \Gamma_t\left(-x\right).$$ 
Then  $\psi_t \mapsto \Gamma_t (-\mathcal{P}_{t+1} (\trans{S_{t+1}} \psi_t))$ is a set-valued function with values in the space $\mathbb{F}$. And equation~\eqref{BellmanB}, and thus~\eqref{Bellman},  can be rewritten as
\begin{align}
%\label{Bellman2}
\begin{split}
\mathcal{P}_{t} \left( v_t \right)  = \; \inf\limits_{\substack{\trans{S_t} \psi_t = v_t,\\ \psi_t \in \Phi_t}} \Gamma_t (-\mathcal{P}_{t+1} (\trans{S_{t+1}} \psi_t)).
\end{split}
\end{align}
Thus, the value function at time $t$ is a one step minimization problem of the mean-risk function $\Gamma_t$, applied to the value function at time $t+1$. This provides an interpretation of~\eqref{Bellman} in total analogy to the scalar case~\eqref{Bell}: instead of a conditional expectation the corresponding mean-risk function $\Gamma_t$ is applied to the value function one time ahead, and the infimum over all possible controls $\psi_t$ is taken. This supports our interpretation of~\eqref{Bellman} as a Bellman equation.

Computational implementations and challenges will be discussed in the next section. Furthermore, we will see in Subsections~\ref{sec_TradingStrategy1} and~\ref{EC_TradingStrategy2}, how the upper images $\mathcal{P}_{t} \left( v_t\right)$ of~\ref{ProbRRt} computed backwards in time can be used to compute an optimal trading strategy of the original dynamic mean risk problem~\ref{Probt} forward in time on the realized path.

%%%%%%%%%%%%%%%%%%%%%%%%%%%%%%%%%%%%%%%%%%%%%%%%%%
\section{Implementing the backward recursion}%%%%%%%%%%%%%%%%%%%%%%%%%
%%%%%%%%%%%%%%%%%%%%%%%%%%%%%%%%%%%%%%%%%%%%%%%%%%
\label{sec_implementation}

In the previous section the recursive relation~\eqref{Bellman}  representing a set-valued Bellman's principle for the mean-risk problem was derived. The next natural step is to use~\eqref{Bellman} and the corresponding recursive vector optimization problem~\ref{ProbRRt} to solve the mean-risk problem backwards in time. In this section we discuss some related challenges.

In theory the recursive problem~\ref{ProbRRt} provides a way to solve the mean-risk problem via backward recursion. However, an application of it in practice is in general not straightforward for the following reason. To solve problem~\ref{ProbRRt}, the time $t+1$ upper image $\mathcal{P}_{t+1}$ needs to be available for any wealth $v_{t+1} = \trans{S_{t+1}} \psi_t$. In general there could be infinitely many values $v_{t+1}$, so infinitely many problems $\tilde{D}_{t+1}$ would be needed to be solved. 
In the scalar case, this does not pose a problem as long as the recursive problem can be solved analytically. Then the solution can be given as a function of the wealth. However, for a VOP an analytic solution is out of reach in general. However, in certain special cases, this problem can be addressed in an easy manner also for VOPs.

For example, the issue would disappear if it was possible to scale the upper images and thus the efficient frontiers for different wealths. This is possible if the risk measure can be scaled. i.e., if the risk measure is coherent. The mean-risk with this additional assumption is studied in this section.

\subsection{The case of a coherent risk measure}
\label{sec_coherent}
The following Lemma provides desirable scaling properties of the mean-risk problem with coherent risk measure. To scale feasible trading strategies, the sets $\Phi_s$ need to be cones.
\begin{restatable}{lemma}{lemmacoherent}
\label{lemma_coherent}
Aside from Assumption~\ref{Assumption1} let the risk measure $(\rho_t)_{t = 0, 1, \dots, T }$ of the investor be coherent, and let each constraint set $\Phi_s$ for $s = 0, \dots , T-1$ be a cone.
Then for any time $t \in \{ 0, 1, \dots, T-1 \}$ and any $v_t > 0$ the following holds:
\begin{enumerate}
\item (Weakly) efficient strategies and (weak) minimizers scale, that is: 
If the portfolio generated by a strategy $(\psi_s)_{s = t, \dots, T-1 }$ is time $t$ (weakly) efficient for initial wealth $\mathbf{1}$ at time $t$, then $(v_t \cdot \psi_s)_{s = t, \dots, T-1 }$ is time $t$ (weakly) efficient for initial wealth $v_t$ at time $t$. 
\item The upper image scales, that is
\begin{align*}
\mathcal{P}_t (v_t)= v_t \cdot \mathcal{P}_{t} (\mathbf{1}). 
\end{align*}
\item (Weak) minimizers of the one-time-step problem scale, that is: If $(\psi_t, x)$ is a (weak) minimizer of~$\tilde{D}_t (\mathbf{1})$, then $(v_t \cdot \psi_t, v_t \cdot x)$ is a (weak) minimizer of ~\ref{ProbRRt}.
\end{enumerate}
\end{restatable}
  
The proof will be given in e-companion to this paper in Section~\ref{EC_proofs}.

Observe that the same scaling principle appears in the standard Markowitz problem. There it is a consequence of a positive homogeneity of the standard deviation as well, which is used to measure the risk. 

A corresponding version of the Lemma~\ref{lemma_coherent} could be proven for negative wealth $v_t <0$ and~$-\mathbf{1}$.  For a general $\mathcal{F}_t$-measurable investment $v_t$, the locality property would enable to scale the strategies and upper image individually in each node. This would, however, complicate the implementation of the problem as the future value of the portfolio depends on the position that is taken, which is a variable of the problem. 

Here we concentrate on a particular case of conical sets $\Phi_s$, the short-selling constraints. These not only simplify the implementation of the problem, but also lead to additional properties studied below. Assumption~\ref{Assumption3} lists all the assumtptions, which will from now on be added to the setting of Assumption~\ref{Assumption1}.
\begin{assumption}
\label{Assumption3} $\;$ 
\begin{enumerate}
\item The risk measure $(\rho_t)_{t = 0, 1, \dots, T }$ of the investor is coherent.
\item Short-selling constraints $\psi_s \geq 0$ for $s = 0, \dots , T-1$ are imposed.
\item The prices are positive, i.e., $S_s > 0$  for $s = 0, \dots , T$, and the investor starts with a positive wealth $v_0 > 0$. 
\end{enumerate}
\end{assumption}
A direct consequence of Assumption~\ref{Assumption3} is a positive value of the portfolio $v_t > 0$ at all times $t$. Then the scaling property of the upper image in Lemma~\ref{lemma_coherent} can be directly used within the recursive problem~\ref{ProbRRt}, which is equivalent to
\begin{align}
\label{Prob_coherent}
\begin{split}
\min\limits_{\psi_t, x} \;\; & 
\begin{pmatrix}
 -\mathbb{E}_t (  -x_1  ) \\
 \;\;\; \rho_t (  -x_2   )
\end{pmatrix} \text{ w.r.t. } \leq_{L_t (\mathbb{R}^2_+)}  \\
\text{s.t. } \;\, & \;\; \trans{S_t} \psi_t = v_t, \\
& \;\; \psi_t \geq 0, \\
& \;\; 
\begin{pmatrix}
x_1  \\ x_2
\end{pmatrix} \in  \left( \trans{S_{t+1}} \psi_t \right) \cdot \mathcal{P}_{t+1} \left( \mathbf{1} \right).
\end{split}
\end{align}
In this formulation it suffices to solve at each time $t$ only one problem, $\tilde{D}_t (\mathbf{1})$. Solutions and upper images for any other wealth can be recovered from it via scaling. This enables us to formulate Algorithm~\ref{alg_0}, which computes the upper image $\mathcal{P}_0 (v_0)$ via a finite number of recursive VOPs. In practice it might be advantageous to solve the node-wise problems instead, for more details see Section~\ref{EC_node} in the e-companion.

\begin{algorithm}
\caption{Computation of $\mathcal{P}_0 (v_0)$}
\begin{algorithmic}[1] \label{alg_0} 
\STATE
\begin{tabular}{ l l}
  Inputs: 	& A financial market satisfying Assumptions~\ref{Assumption1} and~\ref{Assumption3} \\
  			&  initial wealth $v_0 > 0$. \\
\end{tabular}
\STATE $\mathcal{P}_T (\mathbf{1}) := -\mathbf{1}+L_T (\mathbb{R}^2_+)$
\FOR{$t=T-1, \dots, 0$}
\STATE Use $\mathcal{P}_{t+1} (\mathbf{1})$ to solve problem $\tilde{D}_t (\mathbf{1})$, obtain upper image $\mathcal{P}_t (\mathbf{1})$
\ENDFOR
\STATE Scale the upper image, $\mathcal{P}_0 (v_0) = v_0 \cdot \mathcal{P}_0 (\mathbf{1})$
\STATE Output: $\mathcal{P}_0 (v_0)$ and a sequence of upper images $\mathcal{P}_{T-1} (\mathbf{1}), \dots, \mathcal{P}_{0} (\mathbf{1})$
\end{algorithmic}
\end{algorithm}

In the general setting every efficient portfolio is a minimizer of the mean-risk problem (see Lemma~\ref{lemma_minimizer}), and therefore corresponds to a minimal point of the upper image. To obtain a one-to-one relation between the efficient frontier and the upper image, which is an output of the algorithms solving VOPs, one needs also the other direction. This is provided under Assumption~\ref{Assumption3} in the following lemma.

\begin{restatable}{lemma}{lemmaboundedefficient}
\label{lemma_bounded_efficient}
Under Assumptions~\ref{Assumption1} and~\ref{Assumption3} the mean-risk problem~\ref{Probt} is bounded and all minimal points of the upper image $\mathcal{P}_t (v_t)$ correspond to efficient portfolios.
\end{restatable} 

The proof will be given in e-companion to this paper in Section~\ref{EC_proofs}.

\begin{remark}
\label{remark_minimizer_recursive}
Since the problems~\ref{Probt} and~\ref{ProbRRt} share the same upper image, the results of Lemma~\ref{lemma_bounded_efficient} apply to the recursive problem~\ref{ProbRRt} as well. Consider a minimal element (mean-risk profile) $x_t \in \mathcal{P}_t (v_t)$ and a corresponding efficient portfolio $(\psi_s)_{s=t, \dots, T-1}$. The pair $(\psi_t, \Gamma_{t+1}(V_T(\psi)))$ is then a minimizer of~\ref{ProbRRt} that maps to $x_t$. 
\end{remark}

\subsection{Existence and computation of an efficient portfolio}
\label{sec_TradingStrategy1}

Now we return to the question of time consistency under Assumption~\ref{Assumption3}, and will strengthen the results from Section~\ref{sec_TimeConsistency}. It will be shown that for every efficient mean-risk profile  $ x_0^* \in \mathcal{P}_0 (v_0)$ there exists a portfolio that is efficient at all times, and a method to compute such a portfolio will be proposed.

To understand the issue at hand: So far we know that the family of mean-risk problems $\mathcal{D}$ is time consistent w.r.t. weak minimizers (see Theorem \ref{thm_time_consistency2}). That means if $(\psi_s)_{s = t, \dots, T-1}$ is a weak minimizer of~\ref{Probt}, then the truncated strategy $(\psi_s)_{s = t+1, \dots, T-1}$ is a weak minimizer of $D_{t+1} (\trans{S_{t+1}} \psi_t)$.
In general, an efficient portfolio is not guaranteed to remain efficient at subsequent times (it might just be  weakly efficient in a node-wise sense, see Lemma~\ref{lemma_tc_weaknode}).

However, the setting of Assumption~\ref{Assumption3} together with strong monotonicity of the expectation give us something that is stronger than time consistent w.r.t. weak minimizers, but weaker than time consistent w.r.t. minimizers. Lemma~\ref{lemma_strategie1} will show that for every minimal point $ x_0^* \in \mathcal{P}_0 (v_0)$ there \textit{exists} a trading strategy that is a minimizer and stays a minimizer for all time points (which is good enough for an investor, but different from the notion of time consistent w.r.t. minimizers, which would mean that \textit{all} trading strategies that are minimizers stay minimizers).

Given a minimal point $x_0^* \in \mathcal{P}_0 (v_0)$ the desired efficient portfolio can be found by revisiting the recursive problems $\tilde{D}_t$ forwards in time with a fixed objective value (an element of the upper image). For a wealth $v_t^*$ and a mean-risk profile $x_t^* \in \mathcal{P}_t(v_t^*)$ consider the scalar problem
\begin{align}
\tag{$I_t (v_t^*, x_{t}^*)$}
\label{ProbInduction}
\begin{split}
\min\limits_{\psi_t, x_{t+1}} & \;\; \rho_t \left( -x_{t+1,2} \right) \\
\text{s.t.} \;\;\;\;\;\; & \;\; \trans{S_{t}}  \psi_{t}  = v^*_t, \; \psi_{t} \geq 0, \\
&   \;\; 
\begin{pmatrix}
x_{t+1,1}   \\ x_{t+1,2} 
\end{pmatrix}  \in \left( \trans{S_{t+1}}   \psi_{t} \right) \cdot \mathcal{P}_{t+1} \left( \mathbf{1} \right) ,  \\
& \;\; -\mathbb{E}_{t} \left( -x_{t+1,1}  \right)  \leq x_{t,1}^*.
\end{split}
\end{align}
At time $0$ solving $I_0 (v_0, x_0^*)$ yields a pair $(\psi_0^*, x_1^*)$, a minimizer of problem $\tilde{D}_0 (v_0)$ with objective value $x_0^*$. After taking the position $\psi_0^*$ at $t=0$, the investor's portfolio has the value $v_1^* = \trans{S_1} \psi_0^*$ at $t=1$. The random vector $x_1^*$ is an element of $\mathcal{P}_1 (v_1^*)$, but not necessarily a minimal one - it could be just weakly minimal - and therefore it does not necessarily correspond to an efficient portfolio. We will however show, that the way problem~$I_1 (v_1^*, x_{1}^*)$ %\ref{ProbInduction}
is formulated provides a minimizer of problem $\tilde{D}_1 (v_1^*)$ with an objective value that is \textit{at least as good as} $x_1^*$.

\begin{lemma}
\label{lemma_strategie1}
Let Assumptions~\ref{Assumption1} and~\ref{Assumption3} be satisfied, and let a minimal element $x_0^* \in \mathcal{P}_0 (v_0)$ be chosen. Assume that the problems~\ref{ProbInduction} for $t = 0, \dots T-1$ are iteratively solved,
 where the input for the time $t$ problem is given by a solution $(\psi_{t-1}^*, x_{t}^*)$ of the time $t-1$ problem by setting the wealth to be $v_t^* = \trans{S_t} \psi_{t-1}^*$ with $v_0^*=v_0$.
\begin{enumerate}
\item Then for all $t = 0, \dots T-1$  there exists an optimal solution to problem~\ref{ProbInduction}
and any optimal solution $(\psi_t^*, x_{t+1}^*)$ is a minimizer of $\tilde{D}_{t} (v_t^*)$. 

\item The trading strategy $(\psi_s^*)_{s=0, \dots, T-1}$ obtained by this method is an efficient portfolio at time $0$ for wealth $v_0$. Furthermore, the truncated strategy $(\psi_s^*)_{s=t, \dots, T-1}$  is an efficient portfolio for any time $t = 1, \dots T-1$ for the corresponding wealth $v_t^* = \trans{S_t} \psi_{t-1}^*$.
\end{enumerate}
\end{lemma}
 
The proof will be given in e-companion to this paper in Subsection~\ref{EC_TradingStrategy1}.

Let us now return to the motivation behind this work - the portfolio selection problem of the investor. Ultimately the investor is not only interested in knowing the upper image and the efficient frontiers, but also in finding a trading strategy she needs to follow once she selected an efficient portfolio.
Lemma~\ref{lemma_strategie1} directly provides a way to compute the trading strategy $(\psi_s^*)_{s=0, \dots, T-1}$ the investor needs to follow to obtain the mean-risk profile $x_0^* \in \mathcal{P}_0 (v_0)$. However, in practice the investor only needs to know the strategy along the realized path. The corresponding algorithm as well as a simplification in the polyhedral case are discussed in Subsection~\ref{EC_TradingStrategy2}. 

The results here and the algorithms in Subsection~\ref{EC_TradingStrategy2} work with an efficient mean-risk profile $x_0^*$ as an input, but no restriction is placed on how the investor selects it. At least three possibilities suggest themselves - the investor can specify (a) the desired value of the risk measure (risk budget), (b) the desired expected terminal value, or (c) her (initial) risk aversion. Each of these options correspond to one approach to scalarize the mean-risk problem, and in each case the corresponding minimal element of the upper image can be easily found.

%%%%%%%%%%%%%%%%%%%%%%%%%%%%%%%%%%%%%%%%%%%%%%%%%%%%%%%
\section{Examples}%%%%%%%%%%%%%%%%%%%%%%%%%%%%%%%%%%%%
%%%%%%%%%%%%%%%%%%%%%%%%%%%%%%%%%%%%%%%%%%%%%%%%%%%%%%%
The results of Section~\ref{sec_implementation} are now illustrated with two examples. The scalable setting of Assumptions~\ref{Assumption1} and~\ref{Assumption3} is, for convenience, combined with the additional assumption of independent and identically distributed returns.
One can easily verify that in that case the upper images for a given time $t$ are identical in each node, conditionally on the same wealth being available. This simplifies the computations as it suffices to solve at each time $t$ only one node-wise optimization problem for wealth $v_t = 1$.

In both of the discussed examples the risk is assessed by the Conditional Value at Risk. The dynamic Conditional Value at Risk is not a time consistent risk measure, however its recursive version, which is utilized here, does have the property of time consistency, for details see~\cite{CheriditoETAL09}. Furthermore, its polyhedral character (see~\cite{EichhornETAL05}) enables us to reformulate the one-time-step problem~\ref{ProbRRt} as a linear vector optimization problem. Bensolve and Bensolve Tools were used for the calculations, see \cite{CiripoiETAL, LohneETAL16}. Alternatively, also the algorithm of  \cite{RusVanderbei03} could be applied, which was already used to solve large one-period mean-risk problems using real-world data in \cite{RusVanderbei03}.

\begin{example}
\label{ex:1}
Firstly, a binomial market model is considered. Daily trading over a period of ten years, leading to a model with $T = 2\, 500$ time periods, is considered. The parameters of the model were selected to obtain a $4\%$ annual mean return of the stock and no return of the bond, representing the discounted market prices. The Conditional Value at Risk is used at level $\alpha = 1\%$. The upper image and the efficient frontier at the beginning of the ten year period, computed via Algorithm~\ref{alg_0} for wealth $v_0 = 100$, are displayed in Figure~\ref{fig:1}.

\begin{figure}{H}
\center
\includegraphics[width = 0.6\textwidth]{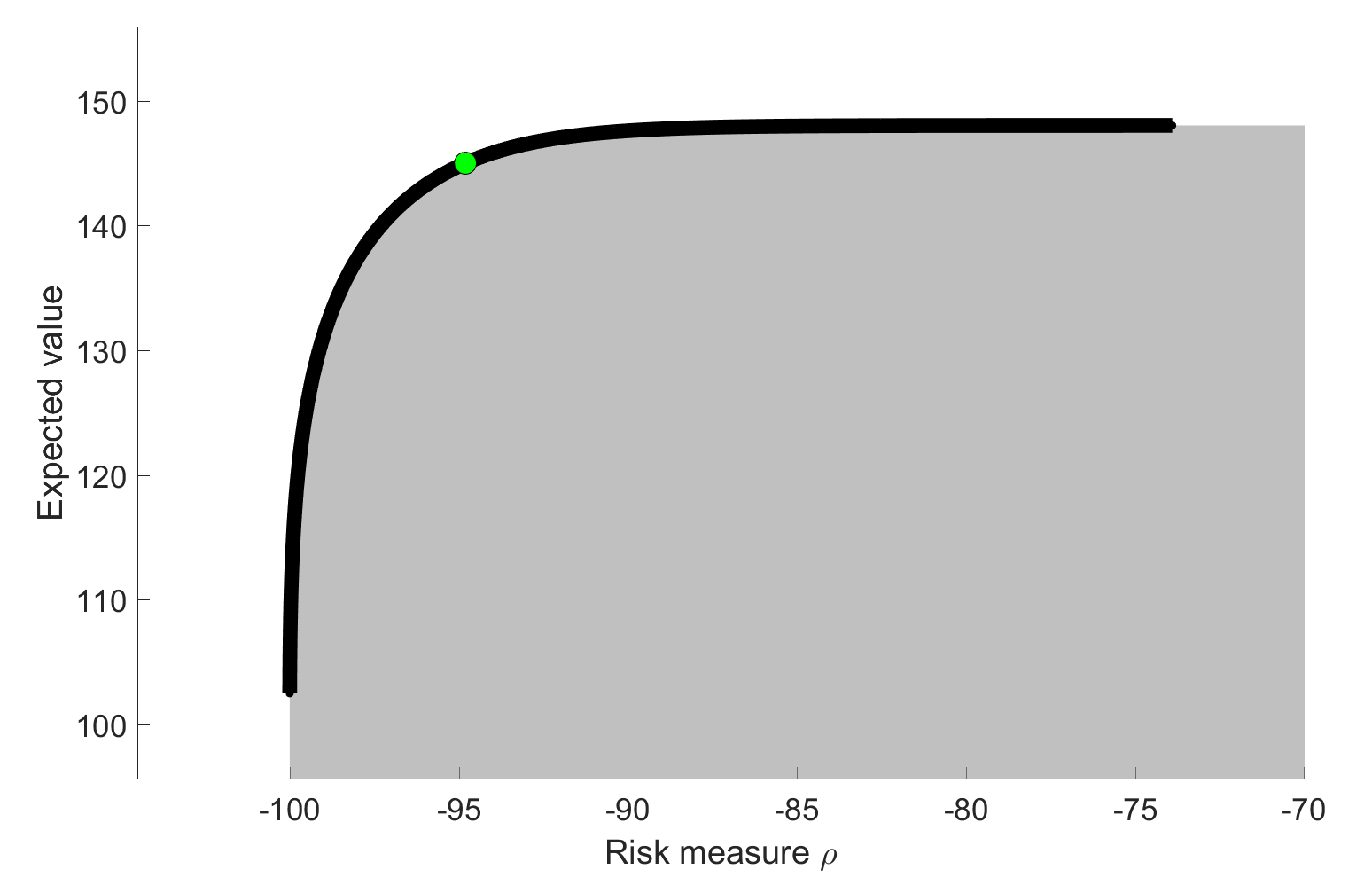}
\caption{Upper image (gray) and efficient frontier (black) in the $(\rho, \mathbb{E})$-plane at time $0$ for a binomial market model with $T=2\, 500$ periods (Example~\ref{ex:1}) for initial wealth $v_0 = 100$. A selected mean-risk profile is highlighted in green.}  \label{fig:1} 
\end{figure}

The obtained  upper image is a polyhedron with $477$ vertices. The efficient frontier contains portfolios with expected terminal values ranging between $102.48$ and $148.02$ and risks between $-100$ and $-73.90$, i.e.~these values are not annualized.  We compare the thus obtained efficient frontier with three popular, but time inconsistent, approaches: a fixed risk aversion, the myopic and the naive (equally weighted) strategy. For a fixed risk aversion $\lambda$ we consider a behavior were at each time the element of the frontier that is optimal for $\lambda$ and a corresponding strategy are found, only to be abandoned at the next time. 
The myopic approach (see e.g.~\cite{Mossin68}) repeatedly considers the problem over a horizon of one period. Because of the simplicity of the binomial model, the  myopic problem leads to corner solutions of either full investment in the stock or full investment in the bond, depending on how the weight between the two objectives is chosen. Table~\ref{tab:1} contains the expected terminal values and the recursive Conditional Value at Risk computed for these strategies over the ten year period. Clearly, neither of them is efficient in the dynamic setting and most of them are so far off the efficient frontier, that we did not depict  them in Figure~\ref{fig:1}. The extreme values of the risk measure are a consequence of the tendency of the $CVaR$ to consider in the binomial model the worst-case scenario only.

\begin{table}[H]
\center
\caption{Mean-risk profiles of time inconsistent strategies - fixed risk aversion (for $\lambda = 0.5$ and $\lambda = 0.9$), myopic (Stock only, Bond only) and the naive one. } \label{tab:1}
\begin{tabular}{ c   c   c  c   c  c }
\hline
						& Fixed $\lambda = 0.5$ 	& Fixed $\lambda = 0.9$	& Stock	 only				& Bond	 only				& Equally  weighted  		\\
%						&					&					&	\\ 
\hline \hline
$-\mathbb{E}_0 (v_T)$	& $-140.75$				& $-104.49$					& $-148.02$			& $-100$		& $-121.67$ 	 	 \\
$CVaR_{1\%, 0} (v_T)$	& $ -0.20$ 				& $-45.10$					& $-0.08$			& $-100$		& $-2.85$					\\ \hline
\end{tabular}
\end{table}

Using Algorithm~\ref{alg_1} (see Subsection~\ref{EC_TradingStrategy2} of the e-companion), a trading strategy can be computed for any selected efficient portfolio. For illustration, a target of an expected terminal value of $145$ was chosen, leading to an efficient portfolio with risk of $-94.8$. This is highlighted on the frontier in the Figure~\ref{fig:1} in green. The trading strategy along one representative path computed via Algorithm~\ref{alg_1} is depicted in Figure~\ref{fig:4}. The trading strategy is represented by the value of the portfolio over time (in orange) and the percentage of this value invested in the risky asset (in blue). 
Additionally, the values of the expectation $\mathbb{E}_t (v_T)$ as well as of the negative of the risk measure $-\rho_t (v_T)$ along the path are provided.  
This allows us to observe the following pattern in the trading strategy. As long as the value of the portfolio value is sufficiently high the stock is strongly preferred. When the value of the portfolio is low and gets closer to the current value of the negative risk, the strategy moves away from the stock towards the bond. Additionally, the polyhedral nature and low dimensionality of the upper images enable to easily compute the weights corresponding to the moving scalarization along the considered path. These weights are also depicted in Figure~\ref{fig:4} (dotted pale blue). In line with the intuition, when the trading strategy is doing badly (i.e., the portfolio value is low and the exposure to the stock is reduced), the weight placed on the risk is increased, and vice versa.

\begin{figure}[H]
\center
\includegraphics[width = 0.6\textwidth]{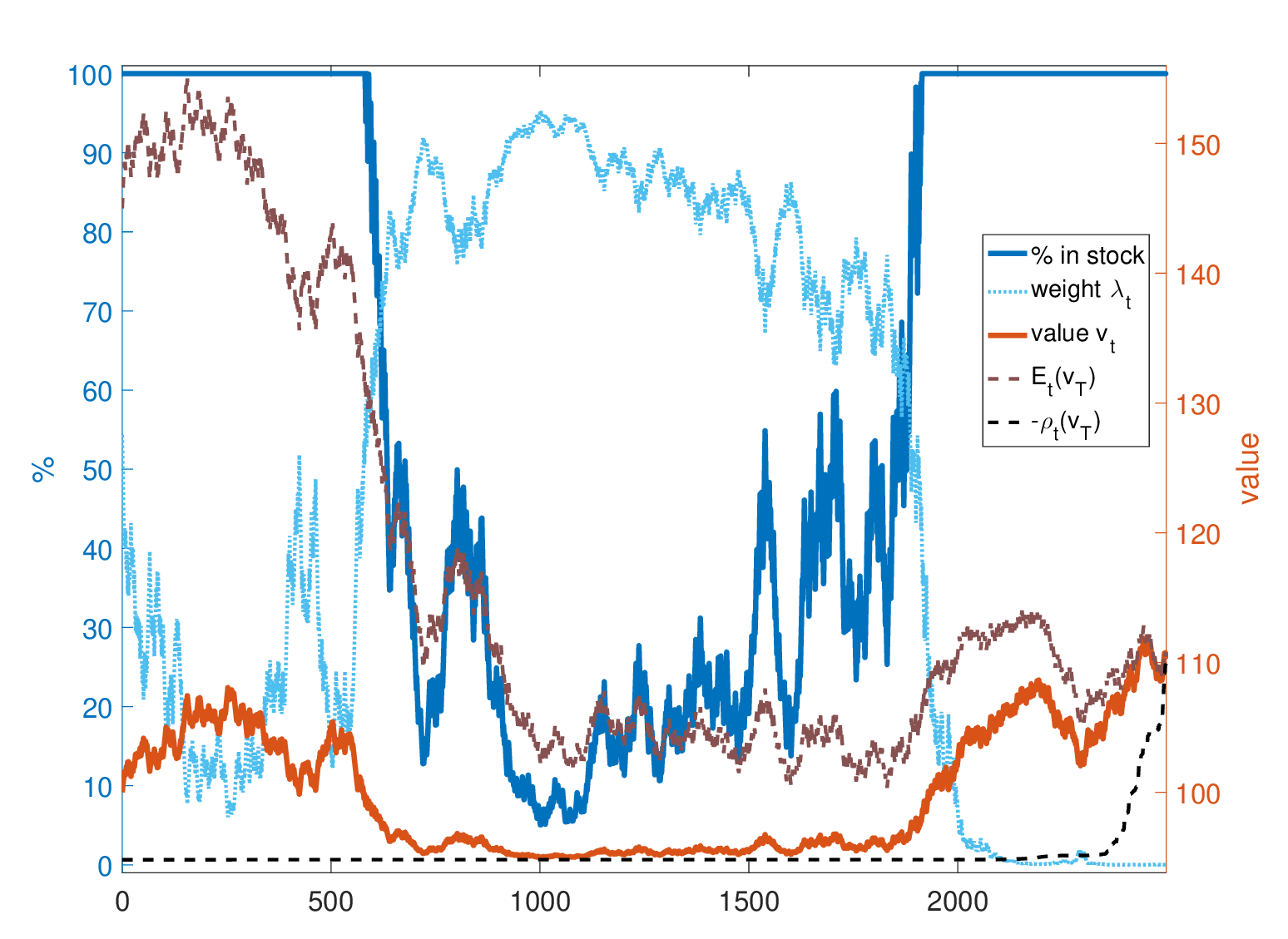}
\caption{Portfolio value (orange) and percentage of that value invested in stock (dark blue) along a selected path in Example~\ref{ex:1}. Additionally, the moving scalarization (dotted pale blue), expectation (dotted brown) and  negative risk  (dotted black) over time are depicted. A scalarization $(1-\lambda_t, \lambda_t)$ is represented by $0 \leq \lambda_t \leq 1$, the risk aversion.} \label{fig:4}
\end{figure}

\end{example}

\vspace{-0.5cm}
\begin{example}
\label{ex:2}
Secondly, a market with multiple assets or asset classes is considered. The first asset is a bond, for the remaining $d=7$, the approach from~\cite{KornETAL09} was used to generate correlated returns. In such a setting, each node of the event tree has $2^d = 128$ successors. Monthly trading over a one year period is used in the example. To compare the effect of the level of the Conditional Value at Risk, the problem was solved for multiple values of $\alpha$. Figure~\ref{fig:5} depicts the obtained efficient frontiers at the initial time for $\alpha = 1\%, 2\%$ and $5\%$.

\begin{figure}[H]
\includegraphics[width = \textwidth]{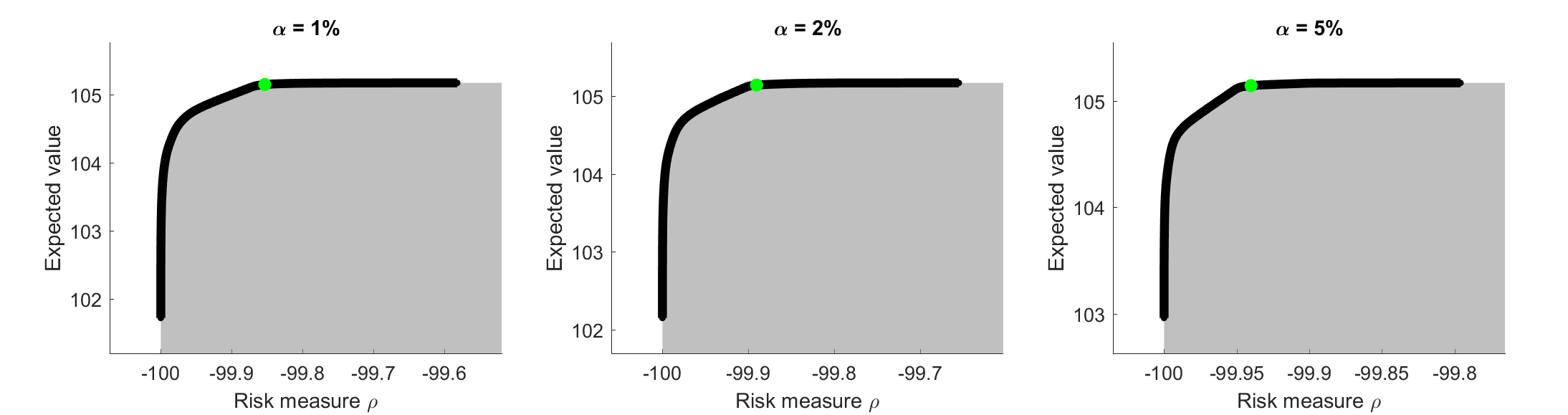}
\caption{Upper images (gray) and efficient frontiers (black) in the $(\rho, \mathbb{E})$-plane at time $0$ in a market with multiple assets with $T=12$ periods (Example~\ref{ex:2}) for initial wealth $v_0 = 100$. The problems were solved for different levels of $\alpha$ of the $CVaR$.  The selected mean-risk profiles are highlighted in green.} \label{fig:5}
\end{figure}

Since in this slightly more complex market model, different levels $\alpha$ of the Conditional Value at Risk lead to different values of the risk measure, the efficient frontiers vary as $\alpha$ changes. The shapes appear similar at different levels $\alpha$ of the risk measure, however the range of the efficient values differs, as does the number of the vertices - the upper image is a polyhedron with $156, 146$ and $107$ vertices for $\alpha = 1\%, 2\%$ and $5\%$, respectively. 
The effects can be observed more drastically in the trading strategies corresponding to efficient portfolios. 
This time a desired portfolio was selected by fixing the risk aversion (scalarization) $\lambda_0$ of the investor and determining an element of the frontier that is optimal for 
$$\min - (1-\lambda_0) \mathbb{E}_0(v_T) + \lambda_0 CVaR_{\alpha, 0} (v_T).$$
Optimal portfolios for $\lambda_0 = 0.5$ were chosen and are highlighted on the frontiers in Figure~\ref{fig:5}. 
The values of their mean-risk profiles are listed in Table~\ref{tab:2} and the trading strategies obtained via Algorithm~\ref{alg_1} along one path are depicted in Figure~\ref{fig:6}.

\begin{table}[H]
\caption{Mean-risk profiles of efficient portfolios highlighted in Figure~\ref{fig:5}.} \label{tab:2}
\begin{tabular}{ c  c   c  c  }
\hline
							& $\; \alpha  = 1\%\;$		& $\; \alpha  = 2\%\;$		& $\; \alpha  = 5\%\;$ \\ \hline \hline
$-\mathbb{E}_0 (v_T)$		& $-105.15$				& $-105.14$				& $-105.14$ \\
$CVaR_{\alpha, 0} (v_T)$	& $-99.85$				& $-99.89$				& $-99.94$ \\ \hline
\end{tabular}
\end{table}

\begin{figure}[H]
\includegraphics[width = \textwidth]{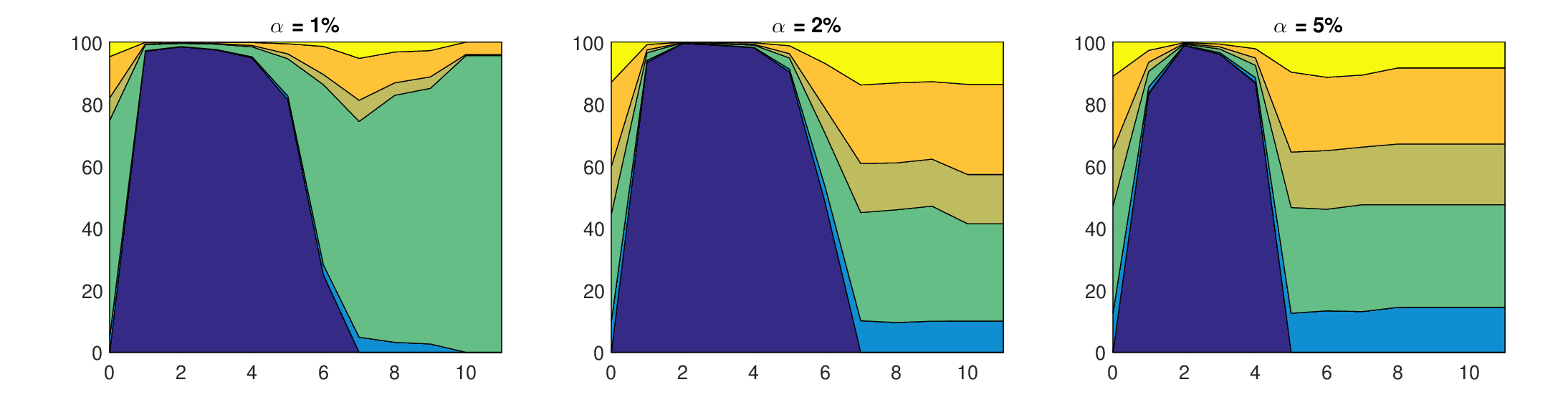} 
\caption{Trading strategies of the selected efficient portfolios along one path in models with $\alpha = 1\%, 2\%$ and $5\%$ (Example~\ref{ex:2}). In dark blue is the bond.} \label{fig:6}
\end{figure}

The most striking feature in Figure~\ref{fig:6} is the tendency to forego diversification in the model with $\alpha = 1\%$. This can be understood as a result of the high weight placed on the worst-case scenario by the dynamic Conditional Value at Risk at this level. As a response, a single asset, which itself has the lowest value of $CVaR$, is disproportionately selected. This behavior is however strongly affected by the parameters of the market model used.

To compare the time consistent dynamic portfolios obtained by the method of this paper with the time inconsistent alternatives, the case of $\alpha = 2\%$ is considered.
It is compared with the myopic portfolio and a strategy arising from a fixed risk aversion, each computed for values $\lambda = 0.5$ and $\lambda = 0.9$, and with the equally weighted portfolio.
While the myopic and the naive strategy have the advantage of an easy computation, Figure~\ref{fig:7} and Table~\ref{tab:3} show them to be inefficient.

\begin{table}[H]
\caption{Mean-risk profiles of the seven trading strategies considered in Example~\ref{ex:2}. } \label{tab:3}
\begin{tabular}{ c   c c   c   c c   c c  }
\hline
						& Dynamic					& Dynamic					& Equally 		& Myopic 			& Myopic				& Fixed  			& Fixed \\
						& $\lambda_0 = 0.5$			& $\lambda_0 = 0.9$			& weighted 		& $\lambda = 0.5$   & $\lambda = 0.9$ 		& $\lambda = 0.5$   & $\lambda = 0.9$        \\ \hline \hline
$-\mathbb{E}_0 (v_T)$	& $-105.14$					& $-104.73$					& $-104.51$ 	& $-105.17$			& $-100$				& $-105.17$			& $-101.23$ \\
$CVaR_{2\%, 0} (v_T)$	& $-99.89$					& $-99.97$					& $-97.86$ 		& $-98.88$			& $-100$ 				& $-98.85$			& $-99.72$ \\ \hline
\end{tabular}
\end{table}

\begin{figure}%[H]
\includegraphics[width = \textwidth]{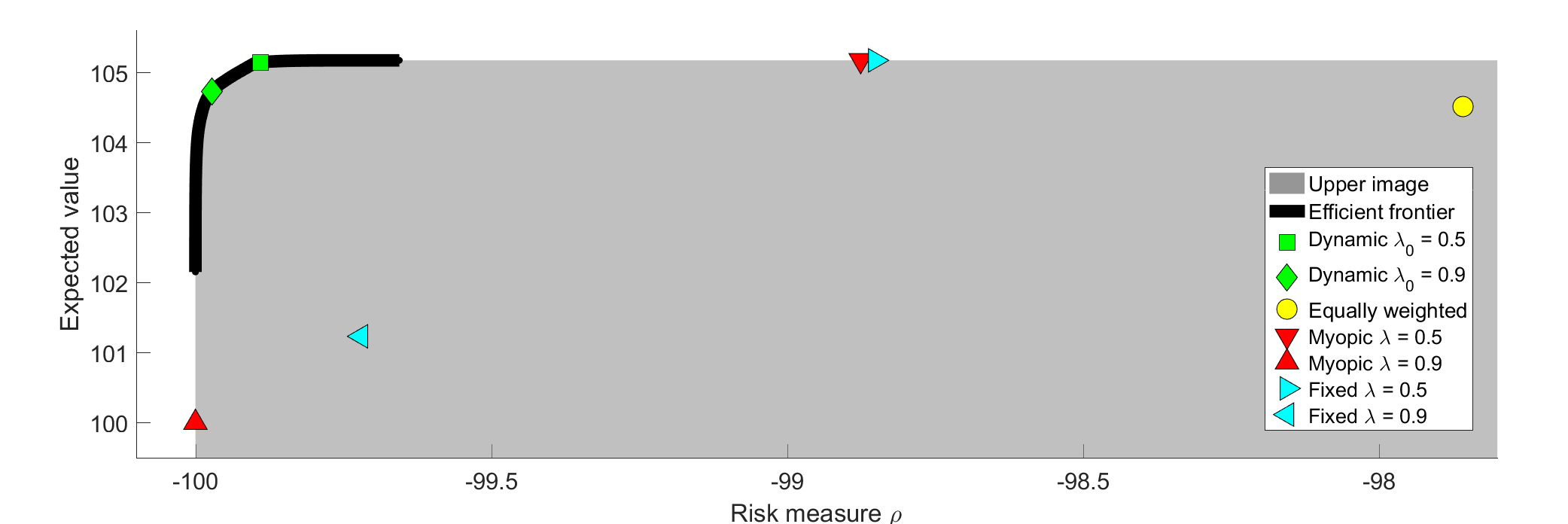} 
\caption{Upper image for $\alpha = 2\%$  in Example~\ref{ex:2} with the seven mean-risk profiles listed in Table~\ref{tab:3}.} \label{fig:7}
\end{figure}

Lastly, Figure~\ref{fig:8} shows the upper image and the efficient frontier of the problem at each time period. All of them are scaled to the corresponding value of the optimal dynamic portfolio along the path depicted in Figure~\ref{fig:6}.   In each step of Algorithm~\ref{alg_1}, the value of the mean-risk profile of the computed portfolio is obtained. These are highlighted on the corresponding frontiers in Figure~\ref{fig:8} in green. Note that these are the optimal (efficient) values rather than the inputs from the previous step of the algorithm, which might be only weakly minimal. Additionally the moving scalarization discussed in Section~\ref{sec_Scalarization} was computed and is included in Figure~\ref{fig:8}. When the mean-risk profile is a vertice of the polyhedral upper image, %a supporting hyperplane and consequently 
the weight is not unique and the obtained interval for $\lambda_t$ is given.

\begin{figure}[H]
\includegraphics[width = \textwidth]{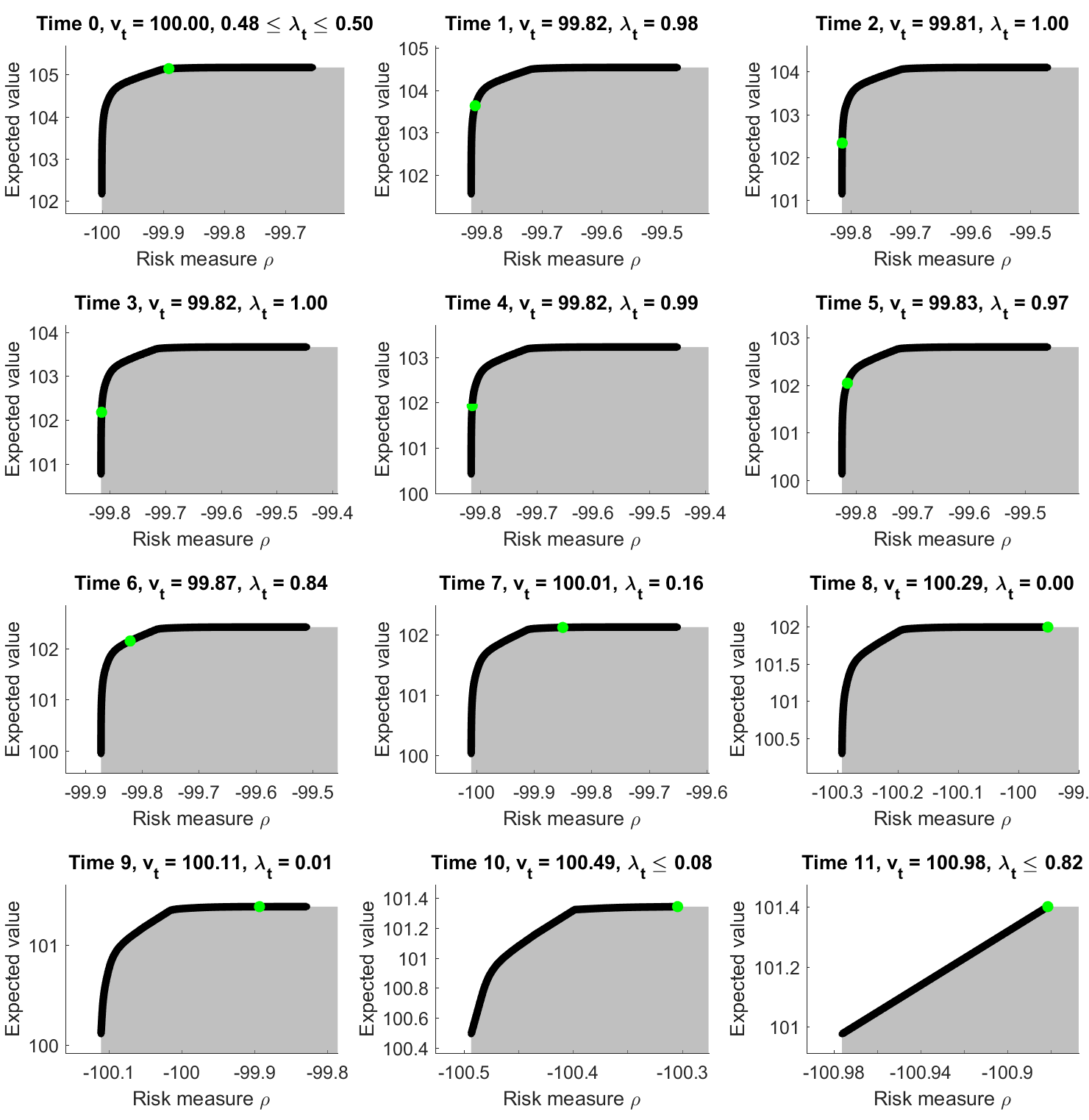}
\caption{Upper images over time for risk measure at level $\alpha=2\%$  in Example~\ref{ex:2}. The upper images are scaled to the value of the portfolio along the path depicted in Figure~\ref{fig:6} starting with the optimal portfolio at initial time with risk aversion $\lambda_0 = 0.5$. Intermediate mean-risk profiles of this portfolio are highlighted in the figures in green. The weights $(1-\lambda_t, \lambda_t)$ corresponding to the moving scalarization along the path are given via the value of $\lambda_t$. } \label{fig:8}
\end{figure}

\end{example}

%%%%%%%%%%%%%%%%%%%%%%%%%%%%%%%%%%%%%%%%%%%%%%%%%
\appendix%%%%%%%%%%%%%%%%%%%%%%%%%%%%%%%%%%%%%%%
%%%%%%%%%%%%%%%%%%%%%%%%%%%%%%%%%%%%%%%%%%%%%%%%

\section{Node-wise Approach}
\label{EC_node}

Throughout the paper the notions of feasibility and efficiency, as well as the optimization problems we defined jointly for all atoms in $\Omega_t$. As only one of them will realize at time $t$, let us now consider the matter from the point of view of a specific node $\omega_t \in \Omega_t$. The definition of efficiency (see Definition~\ref{def_efficient}) can be restricted to the node of interest: 
\begin{definition}
\label{def_efficient_EC}
Under Assumption~\ref{Assumption1}, a feasible portfolio $\psi \in \Psi_t (v_t)$ is \textbf{ time $\mathbf{t}$ efficient for initial wealth $\mathbf{v_t}$ at node $\mathbf{\omega_t \in \Omega_t}$} if, and only if, there exists no other feasible portfolio $\phi \in \Psi_t (v_t)$, such that
\begin{align*}
%\label{def_eff_ineq}
\Gamma_t(V_T(\phi))(\omega_t) \leq_{\mathbb{R}^2_+} \Gamma_t(V_T(\psi))(\omega_t) \;\text{ and }\; \Gamma_t(V_T(\phi))(\omega_t) \neq \Gamma_t(V_T(\psi))(\omega_t).
\end{align*}
It is \textbf{weakly efficient} if the above inequality is strict in both components.
\end{definition}

Clearly, a portfolio that is (jointly) efficient in the sense of Definition~\ref{def_efficient} is also efficient at every node. The reverse is also true.
Locality of the mean-risk function $\Gamma_t(V_T(\cdot))$ (see Lemma~\ref{lemma_Gamma_P} and linearity of $V_T$) and conditionally convexity of  the feasible set $\Psi_t(v_t)$ (see Lemma~\ref{lemma_convexPt}) enable to combine portfolios  that are efficient for individual nodes into a portfolio that is jointly efficient. However, this is not the case for weakly efficient portfolios. Node-wise weak efficiency leads to (joint) weak efficiency in the sense of the above definition, but the reverse is in general not true.

Similarly to efficiency, the mean-risk optimization problem can be formulated in a node-wise fashion -
the objective will be the value of the mean-risk function $\Gamma_t(V_T(\cdot))$ at the given node $\omega_t \in \Omega_t$:
\begin{align}
\label{Probt_EC}
\tag{$D_t (v_t) (\omega_t)$}
\begin{split}
\min \;\;\; &\Gamma_t (V_T (\psi))(\omega_t) \;\text{ w.r.t. }\; \leq_{\mathbb{R}^2_+} \\
\text{s.t. } \; & \psi \in \Psi_t (v_t).
\end{split}
\end{align}

The feasible set of $D_t (v_t) (\omega_t)$ is the full set $\Psi_t (v_t)$, however, the locality property ensures that only the part of the trading strategy relevant at $\omega_t$ influences the objective value. Results of Lemmas~\ref{lemma_minimizer} and~\ref{lemma_convexPt} can be replicated for their node-wise companions, we do not repeat the proofs.
\begin{lemma}
A feasible portfolio $\psi$ is time $t$ (weakly) efficient for initial wealth $v_t$ at node $\omega_t \in \Omega_t$ if, and only if, is it a (weak) minimizer of \ref{Probt_EC}. Problem~\ref{Probt_EC} is a convex vector optimization problem.
\end{lemma}
Since minimizers correspond to efficient portfolios, minimizers of the node-wise problems can be combined into minimizers of the (joint) mean-risk problem. 
The same argument works also for the upper images.

Similarly, a node-wise version of the recursive problem~\ref{ProbRRt} can be formulated. Since the feasible set $\tilde{\Psi}_t (v_t)$ of~\ref{ProbRRt} is conditionally convex (see Lemma~\ref{lemma_conditionally_convex}) and the objective $\Gamma_t$ is local (see Lemma~\ref{lemma_Gamma_P}), the minimizers of the node-wise problems $\tilde{D}_t (v_t) (\omega_t)$ can be combined into minimizers of~\ref{ProbRRt}. The same argument can be made also for the upper images. The node-wise approach might be preferable for implementation.

Mean-risk problems were shown to be time consistent with respect to weak minimizers in Section~\ref{sec_TimeConsistency}. For efficient portfolios these results, however, only mean that they remain at least weakly efficient as the time passes.
 There is a slightly stronger claim, which can be made: an efficient portfolio is at each subsequent time (at least) weakly efficient \textit{in each node}.
\begin{lemma}
\label{lemma_tc_weaknode}
Let $(\psi_s)_{s = t, \dots, T-1}$ be a minimizer of~\ref{Probt}. Then for any time $u > t$ the strategy $(\psi_s)_{s = u, \dots, T-1}$ is a weak minimizer of $D_u (\trans{S_u} \psi_{u-1}) (\omega_u)$ in every node $\omega_u \in \Omega_u$.
\end{lemma}
\begin{proof}
%\proof{Proof.}
Let $(\psi_s)_{s = t, \dots, T-1}$ be a minimizer of~\ref{Probt} and let $u > t$. Assume by contradiction that there exists $\bar{\omega}_u \in \Omega_u$ such that $(\psi_s)_{s = u, \dots, T-1}$ is not a weak minimizer of $D_u (\trans{S_u} \psi_{u-1}) (\bar{\omega}_u)$. Then there exists some feasible $(\phi_s)_{s = u, \dots, T-1} \in \Psi_u(\trans{S_u} \psi_{u-1})$ such that 
\begin{align} 
\label{eq_13}
\Gamma_u (V_T(\phi))(\bar{\omega}_u) \in \Gamma_u (V_T(\psi))(\bar{\omega}_u) - \interior \mathbb{R}^2_+.
\end{align}
Defining $\phi_s := \psi_s$ for $t \leq s < u$ and $\bar{\phi}_s = I_{\bar{\omega}_u} \phi_s + I_{\Omega \backslash \bar{\omega}_u} \psi_s$ for $s \geq t$, two trading strategies $(\phi_s)_{s = t, \dots, T-1}$ and $(\bar{\phi}_s)_{s = t, \dots, T-1}$ are obtained. They are both feasible for problem~\ref{Probt} by Lemma~\ref{Lemma_recursivePsi} and conditional convexity of the feasible set $\Psi_t(v_t)$ (see Lemma~\ref{lemma_convexPt}), respectively. By \eqref{eq_13} it holds $\Gamma_u (V_T(\bar{\phi})) \leq \Gamma_u (V_T(\psi))$ and $\mathbb{P} \left(-\mathbb{E}_u(v_t^{\bar{\phi}}) < -\mathbb{E}_u(v_t^{\psi}) \right) > 0$. By monotonicity and a repeated application of recursiveness one obtains
\begin{align*}
\Gamma_t (V_T(\bar{\phi})) = \Gamma_t \left( -\Gamma_u (V_T(\bar{\phi})) \right) \leq \Gamma_t \left( -\Gamma_u (V_T(\psi)) \right) =  \Gamma_t (V_T(\psi)).
\end{align*}
By strict monotonicity of the expectation one obtains
\begin{align*}
\mathbb{P} \left(-\mathbb{E}_t(v_t^{\bar{\phi}}) < -\mathbb{E}_t(v_t^{\psi}) \right) > 0.
\end{align*}
Together this is a contradiction to $(\psi_s)_{s = t, \dots, T-1}$ being a minimizer of~\ref{Probt}.
%\Halmos \endproof
\end{proof}

\subsection{Existence of an efficient portfolio}
\label{EC_TradingStrategy1}
In Section~\ref{sec_TradingStrategy1}, an efficient portfolio $(\psi_s^*)_{s=0, \dots, T-1}$ corresponding to a selected minimal point $ x_0^* \in \mathcal{P}_0 (v_0)$ is found by solving the scalar problems~\ref{ProbInduction}. These problems are obtained from the recursive problems~\ref{ProbRRt} by setting a constraint on the expectation component and using the risk component as the objective. 
 One can easily verify (compare to Lemma~\ref{lemma_conditionally_convex}) that the problems~\ref{ProbInduction} are convex optimization problems. Just as the vector optimization problems, also these scalar problems can be considered for a specific node $\omega_t \in \Omega_t$:
\begin{align}
\tag{$I_t (v_t^*, x_{t}^*) (\omega_t)$}
\label{ProbInduction_omega}
\begin{split}
\min\limits_{\psi_t(\omega_t), x_{t+1}(\omega_t)} & \;\; \rho_t \left( -x_{t+1,2} \right) \left( \omega_t \right) \\
\text{s.t.} \;\;\;\;\;\; & \;\; \trans{S_{t}} (\omega_t)  \psi_{t}(\omega_t)  = v^*_t (\omega_t), \\
& \;\; \psi_{t} (\omega_t) \geq 0, \\
&   \;\; 
\begin{pmatrix}
x_{t+1,1} (\omega_{t+1})  \\ x_{t+1,2} (\omega_{t+1})
\end{pmatrix}  \in \left( \trans{S_{t+1}} (\omega_{t+1})  \psi_{t}(\omega_t) \right) \cdot \mathcal{P}_{t+1} \left( \mathbf{1} \right) ( \omega_{t+1} ),  \\
&\;\; \forall \omega_{t+1} \in \succ (\omega_t), \\
& \;\; -\mathbb{E}_{t} \left( -x_{t+1,1}  \right) \left( \omega_t \right) \leq x_{t,1}^* (\omega_t),
\end{split}
\end{align}
where we denote for any atom $\omega_t \in \Omega_t$ the set of its successor nodes by $$\succ(\omega_t) := \{ \omega_{t+1} \in \Omega_{t+1}: \; \omega_{t+1} \subseteq \omega_t \}.$$ 
As the risk measure and the expectation are local, only the position $\psi_t (\omega_t)$ and the values $x_{t+1} (\omega_{t+1})$ for $\omega_{t+1} \in \text{succ }(\omega_t)$ are needed as variables of the problem.

The proof of the main result, Lemma~\ref{lemma_strategie1}, uses the following observation, which can be understood as a version of Lemma~\ref{lemma_tc_weaknode} for the one-time-step problem.
\begin{lemma}
\label{lemma_tc_one_step}
Under Assumption~\ref{Assumption1} if $(\psi_t, x_{t+1})$ is a minimizer of \ref{ProbRRt}, then the random vector $x_{t+1}$ is in any node $\omega_{t+1} \in \Omega_{t+1}$ an element minimal in the expectation component of the corresponding upper image  $\mathcal{P}_{t+1} \left( \trans{S_{t+1}} \psi_t \right) (\omega_{t+1})$, that is
\begin{align*}
\left( x_{t+1} (\omega_{t+1}) - \mathbb{R}_{++}\times\mathbb{R}_+  \right) \cap \mathcal{P}_{t+1} \left( \trans{S_{t+1}} \psi_t \right) (\omega_{t+1}) = \emptyset.
\end{align*}
This also implies that $x_{t+1} (\omega_{t+1})$ is a weakly minimal element of $\mathcal{P}_{t+1} \left( \trans{S_{t+1}} \psi_t \right) (\omega_{t+1})$ for every $\omega_{t+1} \in \Omega_{t+1}$ and that the random vector $x_{t+1}$ is a weakly minimal element of $\mathcal{P}_{t+1} \left( \trans{S_{t+1}} \psi_t \right)$.
\end{lemma} 

\begin{proof}
%\proof{Proof.} 
Assume by contradiction that there exists $\bar{\omega}_{t+1} \in \Omega_{t+1}$ such that $\left( x_{t+1} (\bar{\omega}_{t+1}) - \mathbb{R}_{++}\times\mathbb{R}_+  \right) \cap \mathcal{P}_{t+1} \left( \trans{S_{t+1}} \psi_t \right) (\omega_{t+1}) \neq \emptyset$ and denote $y(\bar{\omega}_{t+1})$ an element from this intersection. Define $y(\omega_{t+1}) := x_{t+1} (\omega_{t+1})$ for $\omega_{t+1} \neq \bar{\omega}_{t+1}$. Then $y \in \mathcal{P}_{t+1} \left( \trans{S_{t+1}} \psi_t \right)$ and the pair $(\psi_t, y)$ is feasible for problem \ref{ProbRRt}. By strict monotonicity of the conditional expectation and monotonicity of the risk measure it follows that $\Gamma_t (-y) \leq \Gamma_t (-x_{t+1})$ and $\Gamma_t (-y) \neq \Gamma_t (-x_{t+1})$, which contradicts the assumption that $(\psi_t, x_{t+1})$ is a minimizer.
%\Halmos \endproof
\end{proof}

\begin{proof}[Proof of Lemma~\ref{lemma_strategie1}.]
%\proof{Proof of Lemma~\ref{lemma_strategie1}.}
\begin{enumerate}
\item
The claim will be proven iteratively starting with time $0$. By Remark~\ref{remark_minimizer_recursive}, to a minimal element $x_0^* \in \mathcal{P}_0 (v_0)$ corresponds some minimizer $(\psi_0, x_1)$ of problem $\tilde{D}_0 (v_0)$ with objective value $\Gamma_0 (-x_1) = x_0^*$. One easily sees that $(\psi_0, x_1)$ is feasible also for problem~$I_0 (v_0, x_0^*)$. Assume that $(\psi_0, x_1)$ is not an optimal solution of~$I_0 (v_0, x_0^*)$, then there exists a feasible $(\phi_0, y_1)$ with $\rho_0 (-y_{1,2}) < \rho_0(-x_{1,2})$. Since $(\phi_0, y_1)$ would be feasible also for problem $\tilde{D}_0 (v_0)$, this would contradict minimality of $x_0^*$. Therefore, an optimal solution to~$I_0 (v_0, x_0^*)$ exists. The fact that any optimal solution $(\psi_0^*, x_1^*)$ of~$I_0 (v_0, x_0^*)$ is also a minimizer of $\tilde{D}_0 (v_0)$ with $\Gamma_0 (-x_1^*) = x_0^*$ follows from minimality of $x_0^*$. 

Let the result hold for all $s < t$. As the pair $(\psi_{t-1}^*, x_t^*)$ is a minimizer at time $t-1$, Lemma~\ref{lemma_tc_one_step}  implies that at every node $\omega_t \in \Omega_t$ the vector $x_t^* (\omega_t)$ is minimal in the expectation component and a weakly minimal element of $\mathcal{P}_t (v_t^*) (\omega_t)$. Since the upper image $\mathcal{P}_t (v_t^*)$ is closed, there exists a minimal element $z_t^* \in \mathcal{P}_t (v_t^*)$ such that $z_t^*  \leq x_t^* $. From the minimality of $x_t^* (\omega_t)$ in the  expectation component at every node $\omega_t \in \Omega_t$ follows that $z_{t,1}^*  = x_{t,1}^*$.

By Remark~\ref{remark_minimizer_recursive}, to the minimal element $z_t^*  \in \mathcal{P}_t (v_t^*)$ corresponds some minimizer $(\psi_t , x_{t+1} )$ of problem $\tilde{D}_{t} (v_t^*)$. By the same arguments as for time $0$ one can show that $(\psi_t , x_{t+1}  )$ is feasible and an optimal solution of~\ref{ProbInduction}, as well as that for any optimal solution $(\psi_t^* , x_{t+1}^*)$ of~\ref{ProbInduction} it holds $\Gamma_t (-x_{t+1}^*)  = z_t^*$.  Note that only the expectation component $x_{t,1}^* = z_{t,1}^* $, but not the risk component $x_{t,2}^*$ is an input of problem~\ref{ProbInduction}. Therefore, any optimal solution of~\ref{ProbInduction} is a minimizer of $\tilde{D}_{t} (v_t^*) $. 

\item Let a sequence $(\psi_0^*, x_1^*), \dots, (\psi_{T-1}^*, x_T^*)$ of minimizers of the recursive problems~$\{\tilde{D}_t (v_t^*)\}_{t=0, \dots, T-1}$ be given by the above method, where the corresponding wealths are given by $v_t^* = \trans{S_t} \psi_{t-1}^*$ for $t > 0$. 
From the first part of this proof it follows that at any time $t=0, \dots, T-1$ it holds 
\begin{align}
\label{eq_10}
\Gamma_t (-x_{t+1}^*) \leq x_t^*,
\end{align}
and each $\Gamma_t (-x_{t+1}^*)$ is minimal in its corresponding upper image. Now we follow a backward recursion to show that $(\psi_s^*)_{s=t, \dots, T-1}$ is an efficient portfolio for wealth $v_t^*$ for each time $t = T-1, \dots, 0$. 
Consider time $T-1$. We want to show that the truncated strategy $\psi_{T-1}^*$ is efficient at time $T-1$ for wealth $v_{T-1}^*$. Recall, that the terminal wealth function $V_T$ was defined in~\eqref{terminalwealth} by 
\begin{align*}
V_T (\psi^*) = \begin{pmatrix}
v_T^* \\ v_T^*
\end{pmatrix} = \begin{pmatrix}
\trans{S_T} \psi_{T-1}^* \\ \trans{S_T} \psi_{T-1}^*.
\end{pmatrix}
\end{align*} 
From \eqref{upperT} it follows that $x_T^* \geq -V_T(\psi^*)$, and therefore by monotonicity and \eqref{eq_10} one obtains
\begin{align}
\label{induc}
\Gamma_{T-1} (V_T (\psi^*)) \leq \Gamma_{T-1} (-x_T^*) \leq x_{T-1}^*.
\end{align} 
Since $\Gamma_{T-1} (-x_T^*)$ is minimal, also $\Gamma_{T-1} (V_T (\psi^*))$ is minimal and the portfolio $\psi_{T-1}^*$ is efficient at time $T-1$ for wealth $v_{T-1}^*$.

We will now show that if $\Gamma_{t+1} (V_T (\psi^*)) \leq \Gamma_{t+1} (-x_{t+2}^*) \leq x_{t+1}^*$ holds at time $t+1$, then the corresponding inequality, that is, $\Gamma_{t} (V_T (\psi^*)) \leq \Gamma_{t} (-x_{t+1}^*) \leq x_{t}^*$, holds also at time $t$. The validity of the inequality at time $T-1$ was established in \eqref{induc}. This will then imply the time $t$ efficiency of $(\psi_s^*)_{s = t, \dots, T-1}$ for wealth $v_t^*$. Thus, let us assume that $\Gamma_{t+1} (V_T (\psi^*)) \leq x_{t+1}^*$ holds.
Then, by monotonicity and recursiveness of $\Gamma_t$ and \eqref{eq_10} it follows that
\begin{align*}
\Gamma_t \left(V_T(\psi^*)\right) = \Gamma_t \left( - \Gamma_{t+1} \left(V_T(\psi^*)\right)\right ) \leq \Gamma_t (-x_{t+1}^*) \leq x_t^*.
\end{align*}
Since $\Gamma_t (-x_{t+1}^*)$ is minimal, the portfolio $(\psi_s^*)_{s = t, \dots, T-1}$ is time $t$ efficient for wealth $v_t^*$.
%\Halmos
\end{enumerate}
%\endproof
\end{proof}

\subsection{Computation of the trading strategy}
\label{EC_TradingStrategy2}
Assume the problems $\tilde{D}_{T-1} (\mathbf{1}), \dots, \tilde{D}_{0} (\mathbf{1})$, or their node-wise counterparts, were solved via Algorithm~\ref{alg_0} and the investor selected a desired minimal element $x_0^* \in\mathcal{P}_0 (v_0)$ representing an efficient mean-risk profile. One could directly utilize the method outlined in Section~\ref{sec_TradingStrategy1} to compute an efficient portfolio $(\psi_s^*)_{s = 0, \dots, T-1}$. However, not only would this be computationally expensive, the investor does not need all this information. Ultimately the investor only needs to know what positions she needs to take in nodes which realize. Algorithm~\ref{alg_2} provides a method for computing the portfolio positions along a realized path. Since the sequence of realized nodes is made available only as time passes, the algorithm needs to be applied in real time.
\begin{algorithm}
\caption{Computing an efficient strategy}
\begin{algorithmic}[1] \label{alg_2} 
\STATE
\begin{tabular}{ l l}
   		& A financial market satisfying Assumptions~\ref{Assumption1} and~\ref{Assumption3}, \\
  		& sequence of upper images $\mathcal{P}_{T-1} (\mathbf{1}), \dots, \mathcal{P}_{0} (\mathbf{1})$ from Alg.~\ref{alg_0}, \\
Inputs: 	& initial wealth $v_0 > 0$, \\
  		& a minimal mean-risk profile $x_0^* \in \mathcal{P}_0 (v_0)$,\\
		& a realized path $\omega_0, \omega_1, \dots, \omega_{T-1}$.\\
\end{tabular}
\STATE At time $t=0$ solve problem~$I_0 (v_0, x_0^*)$ obtaining $\psi_0^*$ and $x_1^*$
\FOR{$t=1, \dots, T-1$}
\STATE Update wealth $v_{t}^*(\omega_{t}) = \trans{S_{t}} (\omega_{t}) \psi_{t-1}^*(\omega_{t-1})$
\STATE Solve problem~\ref{ProbInduction_omega} obtaining $\psi_{t}^* (\omega_t)$ and $x_{t+1}^* (\omega_t)$
\ENDFOR
\STATE Output: a sequence of positions $\psi_0^*, \psi_1^* (\omega_1), \dots, \psi_{T-1}^* (\omega_{T-1})$, which represent an efficient portfolio for $x_0^* \in \mathcal{P}_0 (v_0)$ along the realized path
\end{algorithmic}
\end{algorithm}

Algorithm~\ref{alg_2} can be improved if a polyhedral representation of the upper images is available.
Then, all constraints of problem~\eqref{Prob_coherent} can be formulated as linear (in)equalities.  It will be shown below, that in this case each convex optimization problem~\ref{ProbInduction_omega} can be replaced by few arithmetic operations.

Algorithms such as \cite{RudloffETAL17,HamelLohneRudloff14,Lohne11}  provide for a VOP a polyhedral representation of its upper image as well as a solution - i.e. feasible vectors mapping onto vertices and extreme directions of the upper image. For problem $\tilde{D}_t (\mathbf{1}) (\omega_t)$ we introduce the following notation: Let $A (t, \omega_t)$ denote the matrix of vertices of the upper image $\mathcal{P}_t (\mathbf{1})(\omega_t)$, where each column represents one vertex. Let $B (t, \omega_t)$ represent the matrix of solutions. Without loss of generality we assume that the vertices are ordered increasingly by their first component, the negative expectation. The solutions are ordered correspondingly, i.e., column $B_{i} (t, \omega_t)$ is a solution corresponding to vertex $A_{i} (t, \omega_t)$. Since the problem $\tilde{D}_t (\mathbf{1}) (\omega_t)$ is bounded (see Lemma~\ref{lemma_bounded_efficient}),  all minimal points of $\mathcal{P}_t (\mathbf{1})(\omega_t)$ are convex combinations of its (adjacent) vertices.
The next lemma describes how a solution of $\tilde{D}_t (\mathbf{1}) (\omega_t)$ can be used to find an optimal solution of problem~\ref{ProbInduction_omega}.

\begin{lemma}
\label{lemma_polyh}
Let Assumptions~\ref{Assumption1} and~\ref{Assumption3} be satisfied and fix some time $t$ and node $\omega_t \in \Omega_t$. 
Let the positive wealth $v_t^*(\omega_t)$ be given and a point $x_t^*(\omega_t)$ be minimal in the expectation component of the polyhedral upper image $\mathcal{P}_t (v_t^*)(\omega_t)$. An optimal solution  $(\psi_{t}^*(\omega_{t}), x_{t+1}^* (\omega_{t}))$  of problem~\ref{ProbInduction_omega} can be obtained from a solution $B (t, \omega_t)$ of $\tilde{D}_t (\mathbf{1}) (\omega_t)$ and the vertices $A (t, \omega_t)$ of its upper image by setting
$$
\trans{ (\psi_{t}^*(\omega_{t}), x_{t+1}^* (\omega_{t})) }=v_t^*(\omega_t) \cdot \left( w B_{i}(t, \omega_t) + (1-w) B_{i+1} (t, \omega_t)  \right),
$$ where
\begin{align*}
i := \max \Big\{j: A_{1,j} (t, \omega_t) \leq \frac{x^*_{t,1} (\omega_t)}{v_t^* (\omega_t)} \Big\}
\text{ and }
w := \frac{\frac{x^*_{t,1} (\omega_t)}{v_t^* (\omega_t)} - A_{1,i+1}(t, \omega_t) }{A_{1,i}(t, \omega_t) - A_{1,i+1}(t, \omega_t)}.
\end{align*}
\end{lemma}
\begin{proof}
%\proof{Proof.}
Since upper image is closed there exists a minimal point $z_t^* (\omega_t) \leq x_t^* (\omega_t)$ with $z_{t,1}^* (\omega_t) = x_{t,1}^* (\omega_t)$. In the considered setting the upper image scales (see Lemma~\ref{lemma_coherent}), therefore $\frac{z^*_{t} (\omega_t)}{v_t^* (\omega_t)}$ is a minimal element of $\mathcal{P}_t (\mathbf{1})(\omega_t)$. As the vertices in $A (t, \omega_t)$ are assumed to be ordered, a convex combination $u A_{j} (t, \omega_t) + (1-u) A_{j+1}(t, \omega_t)$ leads to a minimal point for any $j$ and $u \in [0,1]$. As the objective function is convex, the corresponding convex combination of solutions $u B_{j} + (1-u) B_{j+1}(t, \omega_t)$ is a minimizer mapping to this minimal point.

For the above choice of index $i$ and weight $w$ one obtains the minimal point $\frac{z^*_{t} (\omega_t)}{v_t^* (\omega_t)}$. By Lemma~\ref{lemma_coherent} minimizers of the one-time-step problem scale, so the combination
$
v_t^*(\omega_t) \cdot \big( w B_{i}(t, \omega_t) + (1-w) B_{i+1} (t, \omega_t)  \big)
$
is a feasible point mapping onto $z_t^* (\omega_t)$ and therefore the desired optimal solution of~\ref{ProbInduction}.
%\Halmos \endproof
\end{proof}

This result is used to modify Algorithm~\ref{alg_2}. Algorithm~\ref{alg_1} replaces the step of finding a solution of the convex optimization problem with  the arithmetic operations of Lemma~\ref{lemma_polyh}, making the procedure computationally easier. A drawback is that it is then necessary to store the solutions of all problems $\tilde{D}_t (\mathbf{1}) (\omega_t)$.

\begin{algorithm}
\caption{Computing an efficient strategy (polyhedral case)} 
\begin{algorithmic}[1] \label{alg_1} 
\STATE
\begin{tabular}{ l l}
		& A financial market satisfying Assumptions~\ref{Assumption1} and~\ref{Assumption3}, \\
  	 	& vertices $A (t, \omega_t)$ of $\mathcal{P}_t (\mathbf{1}) (\omega_t)$ and solutions $B (t, \omega_t)$ of $\tilde{D}_t (\mathbf{1}) (\omega_t)$,  \\
Inputs:  		&  initial wealth $v_0 > 0$, \\
  		& a minimal mean-risk profile $x_0^* \in \mathcal{P}_0 (v_0)$,\\
		& a realized path $\omega_0, \omega_1, \dots, \omega_{T-1}$.\\
\end{tabular}
 \FOR{$t=0, \dots, T-1$}
\STATE Update wealth $v_{t}^*(\omega_{t}) = \trans{S_{t}} (\omega_{t}) \psi_{t-1}^*(\omega_{t-1})$ if $t \neq 0$, otherwise $v_0^* = v_0$
\STATE Set $i := \max \Big\{j: A_{1,j} (t, \omega_t) \leq \frac{x^*_{t,1} (\omega_t)}{v_t^* (\omega_t)} \Big\}$
\STATE Set $w := \frac{\frac{x^*_{t,1} (\omega_t)}{v_t^* (\omega_t)} - A_{1,i+1}(t, \omega_t) }{A_{1,i}(t, \omega_t) - A_{1,i+1}(t, \omega_t)}$
\STATE By setting $\trans{(\psi_{t}^*(\omega_{t}), x_{t+1}^* (\omega_{t}))}:=v_t^*(\omega_t) \cdot \left( w B_{i}(t, \omega_t) + (1-w) B_{i+1} (t, \omega_t)  \right)$ one obtains a solution of problem~\ref{ProbInduction_omega}
\ENDFOR
\STATE Output: a sequence of positions $\psi_0^*, \psi_1^* (\omega_1), \dots, \psi_{T-1}^* (\omega_{T-1})$, which represent an efficient portfolio for $x_0^* \in \mathcal{P}_0 (v_0)$ along the realized path
\end{algorithmic}
\end{algorithm}

\section{Proof of the results from Section~\ref{sec_Bellmanp}}
The aim of this section is to prove the main result, Theorem~\ref{theorem_Bellman}, which means to prove relation~\eqref{Bellman}, which we restate here for the convenience of the reader:
\begin{align}
\tag{\ref{Bellman}}
%\begin{split}
%\mathcal{P}_{t} \left( v_t \right) = \cl \left\lbrace  
%\begin{pmatrix}
% -\mathbb{E}_t (  -x_1  ) \\
% \;\;\; \rho_t (  -x_2   )
%\end{pmatrix} \right.   \Big\vert   \;\; \trans{S_t} \psi_t = v_t, \;\; \psi_t \in \Phi_t, \;\;
%\left. 
%\begin{pmatrix}
%x_1  \\ x_2
%\end{pmatrix} \in \mathcal{P}_{t+1} \left( \trans{S_{t+1}} \psi_t \right) 
%\right\rbrace .
%\end{split}
\begin{split}
\mathcal{P}_{t} \left( v_t \right) = \cl \left\lbrace  
\Gamma_t(-x)  \;\;  \big\vert   \;\; \trans{S_t} \psi_t = v_t, \;\; \psi_t \in \Phi_t, \;\;
x \in \mathcal{P}_{t+1} \left( \trans{S_{t+1}} \psi_t \right) 
\right\rbrace .
\end{split}
\end{align}
In relation~\eqref{Bellman}, all feasible positions $\psi_t$ the investor can hold at time $t$ are considered, as well as all elements of the time $t+1$ upper image corresponding to those positions $\psi_t$. Onto those elements of the time $t+1$ upper image a mean-risk function $\Gamma_t: L_{T} (\mathbb{R}^2) \rightarrow  L_{t} (\mathbb{R}^2)$ 
is applied. The function $\Gamma_t$ has properties similar to a risk measure, which are used in the subsequent proofs. We list them below together with a property of the upper image that will be needed later. Since the ordering cones $L_{t} (\mathbb{R}^2_+), L_{T} (\mathbb{R}^2_+)$ correspond to the natural element-wise orderings in the corresponding spaces, we denote, for convenience and readability, orders generated by them with $\leq$ only.

\begin{lemma}
\label{lemma_Gamma_P} 
The function $\Gamma_t: L_{T} (\mathbb{R}^2) \rightarrow  L_{t} (\mathbb{R}^2)$ has the following properties: for any $X, Y \in L_{T} (\mathbb{R}^2), r \in L_{t} (\mathbb{R}^2)$ and $\alpha \in L_t$, the following holds
\begin{itemize}
\item Conditional translation invariance: $\Gamma_t (X + r) = \Gamma_t (X) - r$,
\item Monotonicity: if $X \leq Y$, then $\Gamma_t (Y) \leq \Gamma_t (X)$,
\item Conditional convexity: $\Gamma_t (\alpha X + (1 - \alpha)Y) \leq  \alpha \Gamma_t (X) + (1 - \alpha) \Gamma_t (Y)$ for $0 \leq \alpha \leq 1$,
\item Recursiveness: $\Gamma_t (X) = \Gamma_t (-\Gamma_{t+1} (X))$,
\item Locality: $\Gamma_t(I_A X) = I_A \Gamma_t (X)$ for any $A \in \mathcal{F}_t$,
\item Continuity: $\lim\limits_{n \to \infty} \Gamma_t (X^n) = \Gamma_t (X)$ when $\lim\limits_{n \to \infty} X^n = X$.
\end{itemize}

An upper image $\mathcal{P}$ of a VOP with a convex ordering cone $C$ satisfies the following monotonicity property: if $p \in \mathcal{P}$ and $p \leq_{C} q$, then $q \in \mathcal{P}$. 
\end{lemma}

\begin{proof}
%\proof{Proof.} 
The properties of $\Gamma_t$ follow from the corresponding properties of the conditional expectation and the risk measure applied component-wise. Convexity of a cone $C$ corresponds to $C + C \subseteq C$. This implies the inclusion $\mathcal{P}+C\subseteq \mathcal{P}$, which is the above stated monotonicity property of $\mathcal P$. 
%\Halmos \endproof
\end{proof}

For the mean-risk problem, time consistency of $\Gamma_t$ specifically means
\begin{align}
\label{eq_5}
\Gamma_t (V_T(\psi)) = \Gamma_t ( -\Gamma_{t+1} (V_T(\psi))),
\end{align}
for any $\psi = (\psi_s)_{s = t, \dots, T-1} \in \Psi_t (v_t)$, where in~\eqref{eq_5} the notation $\psi$ is used once for the portfolio in the time interval $[t, T)$, and once for the portfolio in the time interval $[t+1, T)$. Based on equation~\eqref{eq_5} consider the auxiliary problem 
\begin{align}
\label{ProbRt}
\tag{$\bar{D}_t (v_t)$}
%\begin{split}
%\min\limits_{\psi_t, x} \;\; &
%\begin{pmatrix}
% -\mathbb{E}_t (  -x_1  ) \\
% \;\;\; \rho_t (  -x_2   )
%\end{pmatrix} \text{ w.r.t. } \leq_{L_t (\mathbb{R}^2_+)}  \\
%\text{s.t. } \;\, & \;\; \trans{S_t} \psi_t = v_t, \\
%& \;\; \psi_t \in \Phi_t, \\
%& \;\; 
%\begin{pmatrix}
%x_1  \\ x_2
%\end{pmatrix} \in \Gamma_{t+1} \left( \Psi_{t+1} (\trans{S_{t+1}} \psi_t) \right).
%\end{split}
\begin{split}
\min\limits_{\psi_t, x} \;\; &
\Gamma_t(-x) \text{ w.r.t. } \leq_{L_t (\mathbb{R}^2_+)}  \\
\text{s.t. } \;\, & \;\; \trans{S_t} \psi_t = v_t, \\
& \;\; \psi_t \in \Phi_t, \\
& \;\; 
x \in \Gamma_{t+1} \left( \Psi_{t+1} (\trans{S_{t+1}} \psi_t) \right).
\end{split}
\end{align}
For the problem to be well defined at all time points, including the pre-terminal time $T-1$, we set
\begin{align}
\label{setT}
\Gamma_{T} \left( \Psi_{T} (v_T)  \right) := \left\lbrace \begin{pmatrix} -v_T \\ -v_T \end{pmatrix} \right\rbrace.
\end{align}
The feasible set of the problem~\ref{ProbRt} is denoted by $\bar{\Psi}_t (v_t)$, and its image by $\Gamma_t (\bar{\Psi}_t (v_t))$.
The following lemma shows a close connection between problems~\ref{Probt} and~\ref{ProbRt}.

\begin{lemma}
\label{lemma_recursiveProb}
For any time $t = 0, \dots, T-1$, and for any $\mathcal{F}_t$-measurable investment $v_t$, the mean-risk problem~\ref{Probt} and the auxiliary problem~\ref{ProbRt} share the same image of the feasible set, that is,
\begin{align*}
\Gamma_t (\Psi_t (v_t)) = \Gamma_t (\bar{\Psi}_t (v_t)).
\end{align*}
\end{lemma}
\begin{proof}
%\proof{Proof.} 
Considering~\eqref{setT}, the equivalence at time $T-1$ is straightforward. For all other times~$t$ and investments $v_t$ the equivalence follows from~\eqref{eq_5} and Lemma~\ref{Lemma_recursivePsi}. 
%\Halmos \endproof
\end{proof}

\begin{comment}
%%%%%%%%
\begin{remark}
To derive~\eqref{eq_6}, and to prove the Lemma~\ref{lemma_recursiveProb}, only time consistency of the individual objectives - tower property of expectation and recursiveness of the risk measure - are used. The property of time consistency (w.r.t.  weak minimizers) of the mean-risk problems is not used. In fact, this property could be (redefined and) proven also for a family of problems $\bar{\mathcal{D}} = \{ \bar{D}_t(v_t) \; \vert \; t \in \{0, \dots, T-1\}, \; v_t \text{ is } \mathcal{F}_t\text{-measurable} \}$. Every (weak) minimizers of a vector optimization problem corresponds to a (weakly) minimal element of the image of the feasible set. Therefore through their common image of the feasible set, a binary relation can be established between the (weak) minimizers of the problem~\ref{ProbRt} and~\ref{Probt}. The issue of recovering a corresponding trading strategy will be addressed in the subsection ~\ref{sec_TradingStrategy}. 
\end{remark}
\end{comment}
%%%%%%%%

The auxiliary problems~\ref{ProbRt} are already recursive, as each problem~\ref{ProbRt} uses in its constraints the image of the feasible set of its successor problem~$\bar{D}_{t+1}$. However, they will only serve as a stepping stone to prove the recursiveness of problems~\ref{ProbRRt} in relation~\eqref{Bellman}.
There are several reasons for that. Firstly, problems~\ref{ProbRt}, despite being recursive, would not be suitable for practical implementations as the available solvers for VOPs (\cite{HamelLohneRudloff14,LohneETAL14,Lohne11,RudloffETAL17,RusVanderbei03}) provide the user with the upper image, rather than the images of the feasible set. One may note that the only difference to problem~\ref{ProbRRt} is indeed that in the constraints the image of the feasible set is replaced by the upper image. It will be proven in Lemma~\ref{thm_identical_upper_images} that this will not change the upper images and thus neither the solutions nor the efficient points of the problems. The second reason for considering~\ref{ProbRRt} instead of~\ref{ProbRt} is that it is not clear whether problem~\ref{ProbRt} is convex, in particular, if its feasible set $\bar{\Psi}_t (v_t)$ is convex. Enlarging the feasible set  by replacing the image of the feasible set by the upper image will ensure that we obtain indeed a convex VOP. This will be proven in Lemma~\ref{lemma_recursive_CVOP}.

\subsection{Proof of Lemma~\ref{thm_identical_upper_images}}%%%%%%%%%%%%%%%%%%%%%%%
\label{sec proof L1}

Recall now problem~\ref{ProbRRt}, where we kept the objective function $\Gamma_t (-x)$ the same as in problem~\ref{ProbRt} and just enlarged the feasible set by replacing the image of the feasible set $\Gamma_{t+1} ( \Psi_{t+1} (\trans{S_{t+1}} \psi_t) )$ by the upper image $\mathcal{P}_{t+1} \left( \trans{S_{t+1}} \psi_t \right)$. For all $t=1,..., T-1$ consider
\begin{align}
\label{ProbRRt}
\tag{$\tilde{D}_t (v_t)$}
%\begin{split}
%\min\limits_{\psi_t, x} \;\; & 
%\begin{pmatrix}
% -\mathbb{E}_t (  -x_1  ) \\
% \;\;\; \rho_t (  -x_2   )
%\end{pmatrix} \text{ w.r.t. } \leq_{L_t (\mathbb{R}^2_+)}  \\
%\text{s.t. } \;\, & \;\; \trans{S_t} \psi_t = v_t, \\
%& \;\; \psi_t \in \Phi_t, \\
%& \;\; 
%\begin{pmatrix}
%x_1  \\ x_2
%\end{pmatrix} \in \mathcal{P}_{t+1} \left( \trans{S_{t+1}} \psi_t \right),
%\end{split}
\begin{split}
\min\limits_{\psi_t, x} \;\; & 
\Gamma_t(-x) \text{ w.r.t. } \leq_{L_t (\mathbb{R}^2_+)}  \\
\text{s.t. } \;\, & \;\; \trans{S_t} \psi_t = v_t, \\
& \;\; \psi_t \in \Phi_t, \\
& \;\; 
x \in \mathcal{P}_{t+1} \left( \trans{S_{t+1}} \psi_t \right),
\end{split}
\end{align}
where we set
\begin{align*}
%\label{setTT}
\mathcal{P}_T (v_T) := \Gamma_{T} \left( \Psi_{T} (v_T)  \right) + L_T(\mathbb{R}^2_+).
\end{align*}

We will now prove Lemma~\ref{thm_identical_upper_images}
which states that the upper image $\mathcal{P}_{t} \left( v_t\right) $ of the mean-risk problem~\ref{Probt} coincides with the upper image $\tilde{\mathcal{P}}_t \left( v_t \right)$ of problem~\ref{ProbRRt}, i.e.
$\mathcal{P}_{t} \left( v_t\right) = \tilde{\mathcal{P}}_t \left( v_t \right)$.

\begin{proof}[Proof of Lemma~\ref{thm_identical_upper_images}.]
%\proof{Proof of Lemma~\ref{thm_identical_upper_images}.}
By Lemma~\ref{lemma_recursiveProb}, problems~\ref{Probt} and~\ref{ProbRt} have the same images of the feasible sets, and therefore also the same upper images. Thus, proving $\mathcal{P}_{t} \left( v_t\right) = \tilde{\mathcal{P}}_t \left( v_t \right)$ is equivalent to proving $\bar{\mathcal{P}}_{t} \left( v_t\right) = \tilde{\mathcal{P}}_t \left( v_t \right)$.

The objective functions of problems~\ref{ProbRt} and~\ref{ProbRRt} coincide, the two problems differ only in their feasible sets. Clearly, the feasible set $\bar{\Psi}_t (v_t)$ of problem~\ref{ProbRt} is a subset of the feasible set $\tilde{\Psi}_t (v_t)$ of the problem~\ref{ProbRRt}, that is $\bar{\Psi}_t (v_t) \subseteq \tilde{\Psi}_t (v_t)$. As a consequence the same relation holds also for their upper images, that is $\bar{\mathcal{P}}_t (v_t) \subseteq \tilde{\mathcal{P}}_t (v_t)$.

Thus,  it remains only to show that $\bar{\mathcal{P}}_{t} \left( v_t\right) \supseteq \tilde{\mathcal{P}}_t \left( v_t \right)$. This will be done in two steps. Firstly, it will be shown that  $\bar{\mathcal{P}}_{t} \left( v_t\right)  \supseteq \Gamma_t (\tilde{\Psi}_t (v_t))$. In the second step, we will use that to prove $\bar{\mathcal{P}}_{t} \left( v_t\right) \supseteq \tilde{\mathcal{P}}_t \left( v_t \right)$. Since $L_{t} (\mathbb{R}^2_+)$ and $L_{t+1} (\mathbb{R}^2_+)$ are convex cones, all of the upper images used here will have the monotonicity property introduced in Lemma \ref{lemma_Gamma_P}.

Let us now show that $\bar{\mathcal{P}}_{t} \left( v_t\right)  \supseteq \Gamma_t (\tilde{\Psi}_t (v_t))$. Consider an arbitrary point $p \in \Gamma_t (\tilde{\Psi}_t (v_t))$. Thus, to $p$ corresponds some feasible pair $(\psi_t, x) \in \tilde{\Psi}_t (v_t)$, such that $p = \Gamma_t ( -x )$.  Feasibility means in particular that the random vector $x$ belongs to the time $t+1$ upper image $\mathcal{P}_{t+1} \left( \trans{S_{t+1}} \psi_t \right)$. By definition of the upper image, there exists a sequence 
\[
\left\lbrace  x^{(n)} = u^{(n)} + r^{(n)} \right\rbrace_{n=1}^{\infty} \subseteq \Gamma_{t+1}(\Psi_{t+1} (\trans{S_{t+1}} \psi_t)) + L_{t+1} (\mathbb{R}^2_+)
\] 
converging towards $x$,
where $u^{(n)} \in \Gamma_{t+1}(\Psi_{t+1} (\trans{S_{t+1}} \psi_t))$ and $r^{(n)} \in L_{t+1} (\mathbb{R}^2_+)$.
This yields new pairs $(\psi_t, x^{(n)})$, which are feasible for the recursive problem~\ref{ProbRRt}; and $(\psi_t, u^{(n)})$, which are feasible for both~\ref{ProbRt} and~\ref{ProbRRt}. Feasibility in particular means $\Gamma_t ( -u^{(n)} ) \in \bar{\mathcal{P}}_{t} \left( v_t \right)$. Monotonicity of the objective function implies $\Gamma_t ( -u^{(n)} ) \leq \Gamma_t ( -x^{(n)} )$, and combined with the monotonicity property of the upper image one obtains $\Gamma_t ( -x^{(n)} ) \in \bar{\mathcal{P}}_{t} \left( v_t \right)$ for all $n\in\N$.

Finiteness of the underlying probability space ensures continuity of the convex function $\Gamma_t$, see Lemmas~\ref{lemma_Gamma_P} and~\ref{lemma_RM}. The values $\Gamma_t ( -x^{(n)} )$ then converge towards $p = \Gamma_t ( -x )$, and as an upper image is closed this proves that $\bar{\mathcal{P}}_{t} \left( v_t\right)  \supseteq \Gamma_t (\tilde{\Psi}_t (v_t))$.

In the second part of the proof we will show that $\bar{\mathcal{P}}_{t} \left( v_t\right) \supseteq \tilde{\mathcal{P}}_t \left( v_t \right)$. Consider any $p \in \tilde{\mathcal{P}}_{t} \left( v_t \right)$. From the way an upper image is defined, there exists a sequence $\left\lbrace  p^{(n)}  = q^{(n)} + r^{(n)} \right\rbrace_{n=1}^{\infty} \subseteq \Gamma_t ( \tilde{\Psi}_t (v_t) ) + L_t (\mathbb{R}^2_+)$ converging to $p$, where $q^{(n)} \in \Gamma_t ( \tilde{\Psi}_t (v_t) )$ and $r^{(n)} \in L_t (\mathbb{R}^2_+)$. For each index $n$ we know, from the previous part of the proof, that $q^{(n)} \in \bar{\mathcal{P}}_{t} \left( v_t \right)$. Thus, by the monotonicity property of the upper image also $p^{(n)} \in \bar{\mathcal{P}}_{t} \left( v_t \right)$ for all $n\in\N$. Thus, also its limit $p$ belongs to $\bar{\mathcal{P}}_{t} \left( v_t \right)$ as the upper image is closed by definition. 
%\Halmos \endproof
\end{proof}

\subsection{Proof of Theorem~\ref{theorem_Bellman}}%%%%%%%%%%%%%%%%%%%%%%%%%%%
\label{sec proof T1}

We are now ready to prove the recursive form~\eqref{Bellman} of the upper images $\mathcal{P}_{t} \left( v_t \right)$ of the mean-risk problem~\ref{Probt}, respectively the one-time-step problems~\ref{ProbRRt}.

\begin{proof}[Proof of Theorem~\ref{theorem_Bellman}.]
%\proof{Proof of Theorem~\ref{theorem_Bellman}.}
Lemma~\ref{thm_identical_upper_images} establishes an equivalence between the upper images of the mean-risk problems~\ref{Probt} and the upper images of the recursive problems~\ref{ProbRRt}. As a consequence, the upper images of the mean-risk problems have the following recursive form,
\begin{align}
\label{eq_7}
%\mathcal{P}_{t} \left( v_t \right) = \cl\left( \left\lbrace  
%\begin{pmatrix}
% -\mathbb{E}_t (  -x_1  ) \\
% \;\;\; \rho_t (  -x_2   )
%\end{pmatrix} \right. \right.  \Big\vert   \;\; \trans{S_t} \psi_t = v_t, \;\; \psi_t \in \Phi_t, \;\;
%\left. \left.
%\begin{pmatrix}
%x_1  \\ x_2
%\end{pmatrix} \in \mathcal{P}_{t+1} \left( \trans{S_{t+1}} \psi_t \right) 
%\right\rbrace + L_t (\mathbb{R}^2_+) \right) .
\mathcal{P}_{t} \left( v_t \right) = \cl\left( \left\lbrace  
\Gamma_t(-x) \;\;  \big\vert   \;\; \trans{S_t} \psi_t = v_t, \;\; \psi_t \in \Phi_t, \;\;
x \in \mathcal{P}_{t+1} \left( \trans{S_{t+1}} \psi_t \right) 
\right\rbrace + L_t (\mathbb{R}^2_+) \right) .
\end{align}
What remains to be shown is its equality to the right-hand side of~\eqref{Bellman}, that is that the cone can be omitted in~\eqref{eq_7}. 
Thus, one has to show the following
$$\Gamma_t (\tilde{\Psi}_t (v_t)) + L_t (\mathbb{R}^2_+) \subseteq \Gamma_t (\tilde{\Psi}_t (v_t)).$$
Consider an element $p + r$ of the set on the left-hand side, where $p \in \Gamma_t (\tilde{\Psi}_t (v_t))$ and $r  \in L_t (\mathbb{R}^2_+)$. To the random vector $p$ in the set $\Gamma_t (\tilde{\Psi}_t (v_t))$ corresponds some feasible pair $(\psi_t, x ) \in \tilde{\Psi}_t (v_t)$, such that $p = \Gamma_t ( -x )$.
Since the upper image $\mathcal{P}_{t+1} (\trans{S_{t+1}} \psi_t)$ has the monotonicity property, also a pair $(\psi_t, x + r \mathbf{1})$ is feasible, that is $(\psi_t, x + r \mathbf{1}) \in \tilde{\Psi}_t (v_t)$. The translation invariance of $\Gamma_t$ yields 
$$p + r = \Gamma_t(  -x - r \mathbf{1} ) \in \Gamma_t (\tilde{\Psi}_t (v_t)). \; $$
% \endproof
\end{proof}

\subsection{Proof of Lemma~\ref{lemma_recursive_CVOP}}%%%%%%%%%%%%%%%%%%%%%%%%%%%%
\label{sec proof L2}

For computation and implementation purposes convexity of the recursive problem is needed. 

\begin{lemma}
\label{lemma_conditionally_convex}
The feasible set $\tilde{\Psi}_t (v_t)$ of the problem~\ref{ProbRRt} is conditionally convex.
\end{lemma}
\begin{proof}
%\proof{Proof.}
Let $(\psi_t, x )$ and $(\phi_t, u )$ be feasible for problem~\ref{ProbRRt} and let $\alpha \in L_t$ with $ 0 \leq \alpha \leq 1$.
The first two constraints are satisfied for a convex combination $\alpha \psi_t + (1 - \alpha) \phi_t$ by linearity of portfolio value and conditional convexity of the set $\Phi_t$. We need  to show that a convex combination of $x$ and $ u $ will belong to the upper image $\mathcal{P}_{t+1} \left( \trans{S_{t+1}} ( \alpha \psi_t + (1 - \alpha) \phi_t ) \right)$. Let us distinguish two cases.

Firstly, assume that the vectors $x $ and $ u $ are from the sets $\Gamma_{t+1} \left( \Psi_{t+1} ( \; \cdot \; )  \right) + L_{t+1} (\mathbb{R}^2_+)$. Then there must exist trading strategies $(\psi_s)_{s = t+1, \dots, T-1 } $ and $(\phi_s)_{s \in \{ t+1, \dots, T-1 \}} $ feasible for the problems $D_{t+1} (\; \cdot\;)$ such that
\begin{align}
\label{eq_8b}
x \geq \Gamma_{t+1} (V_T(\psi)), \;\; u \geq \Gamma_{t+1} (V_T(\phi))
%\begin{split}
%x_1  & \geq - \mathbb{E}_{t+1} \left( \; \trans{S_T} \psi_{T-1}    \; \right), \;
%u_1    \geq - \mathbb{E}_{t+1} \left( \; \trans{\phi_{T-1}} S_T    \; \right), \\
%x_2  & \geq \;\;\; \rho_{t+1} \left( \; \trans{S_T} \psi_{T-1}    \; \right), \;
%u_2    \geq \;\;\; \rho_{t+1} \left( \; \trans{\phi_{T-1}} S_T    \; \right). 
%\end{split}
\end{align}
By the conditional convexity of $\Gamma_{t+1}$ and linearity of $V_T$,~\eqref{eq_8b} yields
\begin{align}
%\begin{split}
\label{eq_9b}
\alpha x  + (1 - \alpha) u  &\geq \Gamma_{t+1}(V_T( \alpha \psi + (1 - \alpha)\phi ))
%\alpha x_1  + (1 - \alpha) u_1  &\geq  -\mathbb{E}_{t+1} \left( \trans{(\alpha \psi_{T-1} + (1 - \alpha)\phi_{T-1})} S_T   \right), \\
%\alpha x_2  + (1 - \alpha) u_2  &\geq  \;\; \rho_{t+1} \left(  \trans{(\alpha \psi_{T-1} + (1 - \alpha)\phi_{T-1})} S_T   \right).
%\end{split}
\end{align}
Since the self-financing constraints of the non-recursive problems $D_{t+1} (\cdot)$ are linear, and the constraint sets $\Phi_s$ are conditionally convex, the convex combination of the feasible strategies, $(\alpha \psi_{s} + (1 - \alpha)\phi_{s})_{s \in \{ t+1, \dots, T-1 \}}$, is feasible for problem $D_{t+1} \left( \trans{S_{t+1}} \left( \alpha \psi_t + (1 - \alpha) \phi_t \right) \right)$. This combined with the inequalities~\eqref{eq_9b} means that the random vector 
$\alpha x  + (1 - \alpha) u $
is an element of the upper image $\mathcal{P}_{t+1} \left( \trans{S_{t+1}} ( \alpha \psi_t + (1 - \alpha) \phi_t ) \right)$.

Secondly, assume at least one of the vectors $ x $ and $ u  $ is from the boundary of the corresponding upper image. Then there exist sequences of vectors $\{ x^{(n)} \}_{n=1}^{\infty}$ and $\{ u^{(n)} \}_{n=1}^{\infty}$ from the sets $\Gamma_{t+1} \left( \Psi_{t+1} (\cdot )  \right) \, + \, L_{t+1} (\mathbb{R}^2_+)$ converging to $ x $, respectively $ u $. From above we know that every convex combination $\alpha x^{(n)} + (1- \alpha)u^{(n)}$ lies in the upper image for the starting value $\trans{\left( \alpha \psi_t + (1 - \alpha) \phi_t \right)} S_{t+1}$. Since the upper image is a closed set, also the limit of this sequence, 
$\alpha x  + (1 - \alpha) u$, 
belongs to the upper image $\mathcal{P}_{t+1} \left( \trans{S_{t+1}} ( \alpha \psi_t + (1 - \alpha) \phi_t ) \right)$. 
%\Halmos \endproof
\end{proof}

\begin{proof}[Proof of Lemma~\ref{lemma_recursive_CVOP}.]
%\proof{Proof of Lemma~\ref{lemma_recursive_CVOP}.}
Convexity of the feasible set, which follows from Lemma~\ref{lemma_conditionally_convex}, together with convexity of the objective given in Lemma~\ref{lemma_Gamma_P}, establish the convexity of the one-time-step optimization problem~\ref{ProbRRt} .
%\Halmos \endproof
\end{proof}

\section{Proofs of Lemmas}
\label{EC_proofs}
\lemmaconvexPt*
\begin{proof}
%\proof{Proof.}  
It was already argued in Remark~\ref{remark_VOP} that problem~\ref{Probt} is a VOP. What remains to be shown is the conditional convexity of the feasible set, and conditional $L_t (\mathbb{R}^{2}_+)$-convexity of the objective function. 

 Consider any two feasible trading strategies $\psi, \phi \in \Psi_t (v_t)$ and any $\alpha \in L_t$ with $ 0 \leq \alpha \leq 1$. Since the initial condition, as well as the self-financing conditions are linear in the portfolio positions, they hold for a convex combination as well. Since the constraint sets are conditionally convex, $\alpha \psi_s + (1 - \alpha) \phi_s \in \Phi_s$ holds for all $s = t, \dots, T-1$, so the feasible set $\Psi_t(v_t)$ is conditionally convex. 

The terminal value is linear in the trading strategy,
$$v_T^{\alpha \psi + (1 - \alpha) \phi} = \alpha v_T^{\psi} + (1 - \alpha) v_T^{\phi}.$$ 
This, together with the linearity of the conditional expectation and the conditional convexity of the risk measure implies $L_t (\mathbb{R}^2_+)$-convexity of the objective $\Gamma_t(V_T(\cdot))$,
\begin{align}
\label{gamma_convex}
\Gamma_t (V_T(\alpha \psi + (1 - \alpha) \phi)) \leq \alpha \Gamma_t (V_T(\psi)) + (1 - \alpha) \Gamma_t (V_T(\phi)).
\end{align} 
The mean-risk problem is therefore a convex VOP.

Since \eqref{gamma_convex} holds not only for deterministic scalars, but for $\alpha \in L_t$, the objective function $\Gamma_t(V_T(\cdot))$ is conditionally convex and consequently local.
%\Halmos \endproof
\end{proof}

\LemmarecursivePsi*
\begin{proof}
%\proof{Proof.} 
Recalling the definition of the feasible set, the recursive form follows by
\begin{align*}
\Psi_t (v_t)&= \big\{ (\psi_s)_{s=t, \dots, T-1} \; \vert \; \trans{S_t} \psi_t = v_t, \,\,\psi_t \in \Phi_t , \,\,\trans{S_s} \psi_{s-1} = \trans{S_s} \psi_{s},  \,\,\psi_s \in \Phi_s,  \,\,s = t+1, \dots, T-1 \big\} \\
%\Psi_t (v_t) &= \{ (\psi_s)_{s \in \{t, \dots, T-1\}} \; \vert \; \trans{S_t} \psi_t = v_t, \\ 
%& \;\;\;\;\;\;\;\; \trans{\psi_s} R_{s+1} = \trans{\psi_{s+1}} \mathbf{1}, \psi_s \in \Phi_s, s = t, \dots, T-1 \} \\
%&= \{ (\psi_s)_{s \in \{t, \dots, T-1\}} \; \vert \; \trans{S_t} \psi_t = v_t, \psi_t \in \Phi_t, \\ 
%& \;\;\;\;\;\;\;\; \trans{\psi_{t+1}} \mathbf{1} = \trans{S_{t+1}} \psi_t,  \trans{\psi_s} R_{s+1} = \trans{\psi_{s+1}} \mathbf{1}, \psi_s \in \Phi_s, s = t+1, \dots, T-1 \} \\
&= \{ (\psi_s)_{s =t, \dots, T-1} \; \vert \; \trans{S_t} \psi_t = v_t,  \,\,\psi_t \in \Phi_t, \,\, (\psi_s)_{s =t, \dots, T-1} \in \Psi_{t+1} (\trans{S_{t+1}} \psi_t) \}.
\end{align*}
%\Halmos \endproof
\end{proof}

\movingScal*
Before proving the Lemma we recall some results on convex vector optimization problems and their scalarizations from~\cite{Jahn04} and~\cite{Ulus}.
Consider a convex vector optimization problem
\begin{align}
\tag{$C$}
\label{C}
\min\; F(x) \; \text{ w.r.t. } \leq_{\mathbb{R}^q_+} \text{ subject to } x \in S,
\end{align}
where $S \subseteq \mathbb{R}^m$ is a convex set, $F : \mathbb{R}^m \rightarrow \mathbb{R}^q$ is a convex mapping and $\mathbb{R}^q_+$ is the ordering cone. For a weight $w \in \mathbb{R}^q$ its weighted sum scalarization is
\begin{align}
\tag{$C_w$}
\label{C_w}
\min\; \trans{w} F(x) \text{ subject to } x \in S.
\end{align}
An optimal solution of \eqref{C_w} is a weak minimizer of \eqref{C} for every weight $w \in \mathbb{R}^q_+ \backslash \{0\}$ (see Theorem 5.28 in~\cite{Jahn04}, Proposition 4.8 in~\cite{Ulus}). For our results we need the converse, which we recall in the next lemma (see Theorem 5.13 in~\cite{Jahn04}, Theorem 4.9 in~\cite{Ulus}).
\begin{lemma}
\label{lemma_exists_w}
If the feasible set $S$ is non-empty and closed, then for every weak minimizer $\bar{x}$ of~\eqref{C} there exists $w \in \mathbb{R}^q_+ \backslash \{0\}$ such that $\bar{x}$ is an optimal solution to~\eqref{C_w}.
\end{lemma}

\begin{proof}[Proof of Lemma~\ref{moving_Scal}.]
%\proof{Proof of Lemma~\ref{moving_Scal}.}
Under Assumption~\ref{Assumption1}, the mean-risk problem, as well as its node-wise counterpart, fall into the framework of problem~\eqref{C} and have a closed feasible set, see Remark~\ref{remark_VOP} and Lemma~\ref{lemma_convexPt}. 
By Lemma~\ref{lemma_tc_weaknode}, the portfolio $(\psi_s)_{s = t, \dots, T-1}$ is at least a weak minimizer of $D_t (\trans{S_t} \psi_{t-1}) (\omega_t)$ at each time $t$ in each node $\omega_t \in \Omega_t$. Then by Lemma~\ref{lemma_exists_w} there exists a weight $w_t(\omega_t) \in \mathbb{R}^2_+ \backslash \{0\}$ such that $(\psi_s)_{s = t, \dots, T-1}$ is an optimal solution of a $w_t(\omega_t)$-scalarized mean-risk problem. This yields a stochastic sequence of weights  $w_0, w_1 \dots, w_{T-1}$ corresponding to the given efficient portfolio.
%\Halmos \endproof
\end{proof}

\lemmacoherent*
\begin{proof}
%\proof{Proof.} 
We begin with the first assertion. Under the assumptions of Lemma \ref{lemma_coherent}, the feasible set for an investment $\mathbf{1}$ can be scaled into a feasible set for any investment $v_t > 0$, that is
$$\Psi_t (v_t) = v_t \cdot \Psi_t (\mathbf{1}).$$
From the coherency of the risk measure and linearity of the terminal wealth, the conditional positive homogeneity of the vector-valued objective $\Gamma_t (V_T (\cdot))$ follows. As a result the image of the feasible set of problem~\ref{Probt} can also be scaled for any $v_t > 0$, i.e., 
\begin{align*}
\Gamma_t \left( \Psi_t (v_t) \right) = v_t \cdot \Gamma_t \left( \Psi_t (\mathbf{1}) \right). 
\end{align*}
Thus, it holds
\begin{align*}
& \left( \Gamma_t ( V_T( v_t \cdot \psi )) - L_t(\mathbb{R}^2_+) \backslash \{\mathbf{0}\} \right) \cap \Gamma_t \left( \Psi_t (v_t) \right) 
% &= v_t \cdot \left( \Gamma_t ( vT_(\psi)) - L_t(\mathbb{R}^2_+) \backslash \{\mathbf{0}\} \right) \bigcap v_t \cdot \Gamma_t \left( \Psi_t (\mathbf{1}) \right) \\ 
 = v_t \cdot \left( \left( \Gamma_t ( V_T(\psi)) - L_t(\mathbb{R}^2_+) \backslash \{\mathbf{0}\} \right) \cap  \Gamma_t \left( \Psi_t (\mathbf{1}) \right)  \right),
\end{align*}
respectively
\begin{align*}
& \left( \Gamma_t ( V_T (v_t \cdot \psi)) - \interior L_t(\mathbb{R}^2_+) \right) \cap \Gamma_t \left( \Psi_t (v_t) \right) 
% &= v_t \cdot \left( \Gamma_t ( \psi) - \interior L_t(\mathbb{R}^2_+)  \right) \bigcap v_t \cdot \Gamma_t \left( \Psi_t (\mathbf{1}) \right) \\ 
 = v_t \cdot \left( \left( \Gamma_t ( V_T(\psi)) - \interior L_t(\mathbb{R}^2_+)  \right) \cap  \Gamma_t \left( \Psi_t (\mathbf{1}) \right)  \right).
\end{align*}
Therefore $(\psi_s)_{s = t, \dots, T-1 }$ is a (weak) minimizer of $D_t (\mathbf{1})$ if, and only if $(v_t \cdot \psi_s)_{s = t, \dots, T-1 }$ is a (weak) minimizer of~\ref{Probt}.

To prove the second assertion of the lemma, observe that
\begin{align*}
\mathcal{P}_t (v_t) &= \cl \left( \Gamma_t \left( \Psi_t (v_t) \right) + L_t (\mathbb{R}^2_+) \right) =  \cl v_t \cdot \left( \Gamma_t \left( \Psi_t (\mathbf{1}) \right) + L_t (\mathbb{R}^2_+) \right) = v_t \cdot \mathcal{P}_{t} (\mathbf{1}). 
\end{align*}
The feasible set of the one-time-step problem scales, i.e., $\tilde{\Psi}_t (v_t) = v_t \cdot \tilde{\Psi}_t (\mathbf{1})$, since the upper images $\mathcal{P}_{t+1} ( \cdot )$ scale. The third assertion follows by arguments parallel to the first one.
%\Halmos \endproof
\end{proof}

\lemmaboundedefficient*
\begin{proof}
%\proof{Proof.} 
Because of the finiteness of the probability space, there exists an upper bound on the total returns of the assets, define $\bar{R} := \max\limits_{i = 1, \dots, d} \max\limits_{s = 0, \dots, T-1} \max\limits_{\omega \in \Omega} \frac{S_{s+1, i} (\omega)}{S_{s, i} (\omega)} < \infty$. Given an available wealth $v_s > 0$ at time $s$, the short-selling restriction and positive prices imply $\bar{R} v_s \geq v_{s+1} \geq 0$. The same idea applied recursively implies for all feasible trading strategies of \ref{Probt} 
\begin{align}
\label{eq_12}
\bar{R}^{T-t} v_t \geq v_T \geq 0.
\end{align}
Inequality \eqref{eq_12} leads to bounds $0 \geq \rho_t(v_T) \geq -\bar{R}^{T-t} v_t,$ and $ 0 \geq -\mathbb{E} (v_T) \geq -\bar{R}^{T-t} v_t$ on the objectives. The problem is bounded as $\mathcal{P}_t (v_t) \subseteq -\bar{R}^{T-t} v_t \mathbf{1} + L_t(\mathbb{R}^2_+)$. 

The second part of the claim follows from Theorem 2.40 in \cite{Lohne11}, a vectorial version of the Weierstrass theorem. The compactness of the feasible set and the continuity of the objective are sufficient to satisfy the assumptions of that theorem.
 As discussed in Remark~\ref{remark_VOP}, the variable space and objective space of the mean-risk problem \ref{Probt} can be seen as the Euclidian spaces of appropriate dimensions, therefore it remains to prove boundedness of $\Psi_t (v_t)$, as it is closed (see Remark~\ref{remark_VOP}).
Through the same recursion as for the value, one obtains also $\trans{S_s} \psi_s \leq \bar{R}^{s-t} v_t $ for any time $s \in [t, T-1]$. The position in any individual asset $i$ can then be bounded by $\psi_{s,i} \leq \bar{R}^{s-t} \frac{v_t}{S_{s,i}}$. Since the market is modeled on a finite probability space, there exists a strictly positive lower bound on the prices and position $\psi_{s,i}$ can be bounded by a finite constant for any time and any asset. Set $\Psi_t (v_t)$ is bounded. 

Theorem 2.40 of \cite{Lohne11} implies the existence of a solution (in a sense of set optimization) to the mean-risk problem. As a consequence, the set $\Gamma_t \left( \Psi_t (v_t) \right) + L_t (\mathbb{R}^2_+)$ is closed (see Sections~2.4 and~2.5 of \cite{Lohne11}), and therefore every minimal point of the upper image belongs to the set $\Gamma_t \left( \Psi_t (v_t) \right)$ and corresponds to some efficient portfolio. 
%\Halmos \endproof  
\end{proof}

\bibliographystyle{plain}
\bibliography{biblio}

\end{document}